# Do open citations give insights on the qualitative peer-review evaluation in research assessments? An analysis of the Italian National Scientific Qualification

Federica Bologna, Information Science, Cornell University, Ithaca, USA - email: fb265@cornell.edu - ORCID: 0000-0002-3845-8266
Angelo Di Iorio, Department of Computer Science and Engineering, University of Bologna, Bologna, Italy - email: angelo.diiorio@unibo.it - ORCID: 0000-0002-6893-7452
Silvio Peroni, Research Centre for Open Scholarly Metadata, Department of Classical Philology and Italian Studies, University of Bologna, Bologna, Italy - email: silvio.peroni@unibo.it - ORCID: 0000-0003-0530-4305
Francesco Poggi, Institute of Cognitive Sciences and Technologies (ISTC), Italian National Research Council (CNR), Roma, Italy - email: francesco.poggi@cnr.it - ORCID: 0000-0001-6577-5606

## Abstract

In the past, several works have investigated ways for combining quantitative and qualitative methods in research assessment exercises. Indeed, the Italian National Scientific Qualification (NSQ), i.e. the national assessment exercise which aims at deciding whether a scholar can apply to professorial academic positions as Associate Professor and Full Professor, adopts a quantitative and qualitative evaluation process: it makes use of bibliometrics followed by a peer-review process of candidates' CVs. The NSQ divides academic disciplines into two categories, i.e. *citation-based* disciplines (CDs) and *non-citation-based* disciplines (NDs), a division that affects the metrics used for assessing the candidates of that discipline in the first part of the process, which is based on bibliometrics. In this work, we aim at exploring whether citation-based metrics, calculated only considering open bibliographic and citation data, can support the human peer-review of NDs and yield insights on how it is conducted. To understand if and what citation-based (and, possibly, other) metrics provide relevant information, we created a series of machine learning models to replicate the decisions of the NSQ committees. As one of the main outcomes of our study, we noticed that the strength of the citational relationship between the candidate and the commission in charge of assessing his/her CV seems to play a role in the peer-review phase of the NSQ of NDs.

**Acknowledgements**

This research has been supported by the University fund for Research 2020 (FAR) of the University of Modena and Reggio Emilia, and by the European Union's Horizon 2020 research and innovation program under grant agreement No. 101017452.

# Introduction

A key concept of scientometrics is the study and quantification of the scientific impact of a work. In the sociology of science, two competing theories of citing behavior have been developed: the normative theory and the social constructivist view. Following Merton's (1973) sociological theory of science, the *normative theory* affirms that scientists, through the citation of a scientific work, recognize a credit to colleagues whose results they have used. As summarized by Bornmann (2016), "this means that citations represent an intellectual or cognitive influence on their scientific work". A second view on citing behavior is *social constructivist* grounded on theory advanced in the sociology of science (Latour & Woolgar, 1979; Knorr-Cetina, 1981). Constructivists claim that the cognitive content of papers has poor influence on how they are received. This approach questions the base assumptions of normative theory, and so challenges the validity of using citation analysis for evaluation purposes. As stated by Bornmann (2016), "scientific knowledge is socially constructed through the manipulation of political and financial resources and the use of rhetorical devices" (Knorr-Cetina, 1991). The reasons that lead an author to a citation are complex and variously socially constructed, and these motivations have long been acknowledged in the field (Garfield, 1962).

Empirical tests on the validity of the two theoretical approaches were undertaken in Baldi (1998) and White (2004), and investigations on the possibility for a rapprochement of the normative and constructivist theories have been performed by Small (2004).

Several works in the field of scientometrics have investigated and given increasing importance to combined quantitative and qualitative methods of assessment. In 1987, a workshop was organized by John Irvine, Loet Leydesdorff, and Anthony van Raan on "The relations between qualitative theory and scientometric methods in science and technology studies", which resulted in a special issue of Scientometrics (1989). In the introduction, Leydesdorff (1989) claims that there is increasing recognition of the need to integrate qualitative theories with quantitative techniques provided. The various contributions in the issue elucidate the mutual relevance of theory and empirical scientometric work. For example, Luukkonen (1989) advocates for the use of multi-faceted approaches in assessment studies. Moreover, Kranakis and Leydesdorff (1989) put in relation the results of a historical approach and a scientometric analysis (word occurrences and patterns) for the reconstruction of the cognitive organization of teletraffic science.

After this first issue, an increasing number of works have focused on the divide and connection between qualitative and quantitative science studies. For example, Milojevic et al. (2014) find that chapters from both quantitative and qualitative handbooks of Science and Technology Studies (STS) shared similar academic interests; and Wyatt et al. (2017) argue for the integration of quantitative methods within qualitative analyses. One of the most recent contributions to the topic is the recent issue of *Quantitative Science Studies* on "Bridging the divide between qualitative and quantitative science studies" curated by Loet Leydesdorff, Ismael Ràfols, and Staša Milojevic (2020) . Here, Frenken (2020) assembles several theoretical traditions into a single analytical framework and research design by conceptualizing the diffusion of scientific knowledge as different forms of proximity. In addition, Fox (2020) shows that scholarship on gender inequalities in science could benefit from multi-level studies that utilize different theoretical and methodological approaches. Finally, Marres and de Rijcke (2020) present an example of methodology development that combines quantitative approaches with interpretative ones, and propose that the former should be devised according to their context of use.

This study both draws from and expands this discussion in the context of the Italian National Scientific Qualification (NSQ – in Italian, *Abilitazione Scientifica Nazionale*, or ASN). The NSQ is the national assessment exercise which aims at deciding whether a scholar can apply to professorial academic positions as Associate Professor and Full Professor. It consists of a quantitative and qualitative evaluation process, which first makes use of bibliometrics, to filter candidates who pass specific thresholds, then followed by a peer-review process, where candidates' CVs are assessed by a committee (a.k.a. the commission) to get to the final decision.

The NSQ divides academic disciplines into two categories, i.e. *citation-based* disciplines (CDs) and *non-citation-based* disciplines (NDs)[1]. This division affects only the metrics used for assessing the candidates of that discipline in the first part of the process, which is based on bibliometrics. In this work, we aim at exploring whether citation-based metrics can support the human peer-review of NDs and yield insights on how it is conducted. Specifically, our work focuses on citation-centric metrics that are devised to capture the relationship between the candidate of the NSQ and the commission in charge of the decision upon the candidate, by measuring the overlap between their citation networks. Our hypothesis is that citations give insights on the qualitative peer-review evaluation that follows the preliminary assessment, even in the case of NDs. Therefore, citation metrics should act as a proxy for and mimic the outcome of the human evaluation phase. To measure this relation, we tried to analyze such citation data using a computational method that included two machine learning classifiers, namely SVM and Decision Trees. Our aim was also to experiment how AI-aided analysis can support Science of Science research, as suggested in (Fortunato et al, 2018).

Our study makes a further step: it exclusively utilizes open bibliographic data. Indeed, our citation-centric metrics have been calculated using data collected from open access datasets only. This makes it possible for other scholars to reproduce our analysis, and thus, support the quality of our study, and let us investigate the potentialities of such an approach and its repeatability.

[1] According to the NSQ rules and nomenclature, the names used to identify CDs and NDs are *bibliometric disciplines* and *non-bibliometric disciplines* respectively. Naively, this suggests that only disciplines of the first kind are evaluated using bibliographic metrics. However in the first step of the NSQ evaluation all candidates are evaluated using bibliometric information, making this terminology inconsistent with that used by the scientometric community. In order to better contextualize this study and its purposes, we decide to rename these disciplines into CDs and NDs. Indeed, there are two citation-based metrics among those used for assessing the first type of disciplines; whereas, no citation-based metric is used to evaluate the second type of disciplines.

In recent years, many scholars – led by internationally-coordinated efforts such as the Initiative for Open Citations (I4OC, https://i4oc.org) and the Initiative for Open Abstracts (I4OA, https://i4oa.org) – have come forward to advocate for the free accessibility of bibliographic and citation data. This push has resulted in numerous projects, such as COCI (Heibi et al., 2019), OUCI (Cheberkus & Nazarovets, 2019), Unpaywall (Else, 2018), and the extensions of VOSviewer (van Eck & Waltman, 2010) to handle open scholarly metadata and open citations. Previous works have already shown open data's research potential in the study of bibliometric disciplines – e.g. (Bedogni et al., 2022; Bologna et al., 2022; Chudlarský & Dvořák, 2020; Di Iorio et al., 2019; Huang et al., 2019; Martín-Martín et al., 2020; Peroni et al., 2020; Zhu et al., 2019).

Therefore, the research question of this paper can be summarized as follows: *can the citation-centric metrics we computed using open access datasets provide insights on the human evaluation of these disciplines? In particular, what role does the relationship between the candidate and the commission play in the peer-review phase of the NSQ?*

To answer these questions, we ground our analysis on the data of the candidates and commissions that took part in the 2016, 2017, and 2018 terms of the NSQ. A necessary, but not sufficient, condition is that the open access datasets are sufficiently rich to represent the scientific production of these candidates, on top of which we could calculate citation metrics. In fact, we measured how many publications listed in the candidates CVs are found in the open datasets and focused on the two disciplines that were the best covered, among all the NDs and did not belong to the same Scientific Area: *Historical and General Linguistics* and *Mathematical Methods of Economics, Finance and Actuarial Sciences*, having Recruiting Fields[2] 10/G1 (*Antiquities, Philology, Literary Studies, Art History)* and 13/D4 (*Economics and Statistics)* respectively according to the NSQ classification. The coverage was high, as detailed in Section 'Methods and Materials', and we could compute reliable metrics.

We collected the bibliographic metadata and citation data from the following open datasets: Microsoft Academic Graph

[2] Recruiting Fields correspond to specific scientific fields of study (Ministerial Decree 159, 2012).

(https://www.microsoft.com/en-us/research/project/microsoft-academic-graph/) (Wang et al., 2020), OpenAIRE (https://www.openaire.eu/)(Rettberg & Schmidt, 2012), Crossref (https://www.crossref.org/) (Hendricks et al., 2020), and OpenCitations (https://opencitations.net/) (Peroni & Shotton, 2020). In addition, we devised and reused a set of bibliographic and citation-based metrics using a top-down process and combining popular metrics (e.g. bibliographic coupling) to others proposed for the purpose of this study to describe the relationship between the candidates and the commision of the NSQ. Thanks to the NSQ data and the metrics adopted, we verify whether using our metrics as input to a series of computational methods produces similar results to the ones of the human evaluation conducted by NSQ experts.

## The National Scientific Qualification: how it works

In 2011, Italian Law of December 30th 2010 n.240 (L. 240/2010, 2011) introduced extensive changes in the organization and evaluation of higher education institutions and scholars. In particular, it made it mandatory to pass the NSQ in order to apply to academic positions. The NSQ consists of two distinct qualification procedures designed to attest two different levels of scientific maturity of a scholar. The first procedure qualifies the candidate for the role of Full Professor (FP); whereas the second qualifies the candidate for the role of Associate Professor (AP). It is worth mentioning that passing the NSQ does not grant a tenure position. Each university is responsible for creating new positions according to financial and administrative requirements and in compliance with local hiring regulations.

Besides, Ministerial Decree of June 14th 2012 (D.L. 2012, 2012) defines a taxonomy of 184 Recruitment Fields (RF) divided in groups and sorted into 14 different Scientific Areas (SA). SAs correspond to vast academic disciplines, whereas RFs correspond to specific scientific fields of study. Each scholar is assigned to a specific RF which belongs to a single SA. In the taxonomy, RFs are identified by an alphanumeric code in the form AA/GF. AA is a number indicating the SA and it ranges from 1 to 14. G is a single letter identifying the group of RFs. F is a digit indicating the RF. For instance, *Neurology*'s code is *06/D5*, where *06* indicates the SA *Medicine* and *D* indicates the group *Specialized Clinical Medicine* (D.L. 2012, 2012). When applying for the NSQ, scholars can choose to be evaluated for more RFs at a time. Since each RF has its own assessment rules, the candidate may pass the qualification in some fields and not in others.

The first two terms of the NSQ took place in 2012 and 2013. Although L. 240/2010 prescribes that the NSQ must be held at least once a year, the next session took place from 2016 to 2018: with 1 term in 2016, 2 terms in 2017 and 2 terms in 2018. Those who fail the qualification cannot apply again to the same RF in the year following the date of submission of the application. Once acquired, the qualification's certificate lasts for nine years. For each RF, the Ministry of University and Research (MIUR) appoints an evaluation committee (a.k.a. a commission) composed of five full professors responsible for assessing applicants for associate and full professorships. In order to apply to the NSQ, candidates have to submit a curriculum vitae with detailed information about their research accomplishments.

In the preliminary phase of the evaluation process, each candidate's academic expertise is assessed using bibliometrics. These metrics vary depending on whether the candidate has applied to a CD or a ND. Candidates applying to CDs are evaluated using:

- CD_M1: the number of their journal papers;
- CD_M2: the total number of citations received;
- CD_M3: their h-index.

Citation-based metrics are used to evaluate predominantly scientific RFs, for which reliable citation databases exist: all RFs in the first nine SAs (01-09), with the exception of the RFs 08/C1, 08/D1, 08/E1, 08/E2, 08/F1 and the four RFs in Psychology (11/E). While, candidates applying to NDs are evaluated using:

- ND_M1: number of their journal papers and book chapters;
- ND_M2: number of their papers published on *Class A* journals[3];
- ND_M3: number of their published books.

Non-citation-based metrics are applied to predominantly humanistic RFs, for which no sufficiently complete citation database exists: the last five SAs (10-14) with the exceptions described above.

[3] The List of Class A journals is periodically released by the National Agency for the Assessment of Universities and Research (in Italian, *Agenzia Nazionale di Valutazione del sistema Universitario e della Ricerca*, or ANVUR), available at https://www.anvur.it/attivita/classificazione-delle-riviste/classificazione-delle-riviste-ai-fini-dellabilitazione-scientifica-nazionale/elenchi-di-riviste-scientifiche-e-di-classe-a/ (Last accessed 13 January 2021).

In this first step of the evaluation, candidates' metrics are expected to exceed two out of the three thresholds in their RF. Successively, the candidate's maturity is evaluated based on their CV. The aforementioned metrics are computed for each candidate, taking into consideration only publications that are less than 15 years old for candidates to the role of FP and 10 years old for candidates to the role of AP. This process utilizes data retrieved from Scopus and Web of Science and is conducted by ANVUR. ANVUR also sets thresholds for each metric by RF. Normalization based on the scholars' scientific age (the number of years since the first publication) is used to compute the metrics, for both CDs' and NDs' candidates in all of the considered sessions of the NSQ.

# Methods and materials

This section introduces all the methods and material used for our study. All the data gathered and the software developed for our work are available in (Bologna et al., 2021) and in the GitHub repository https://github.com/sosgang/bond.

### Citation Network Analysis and Metrics

The main idea of our work is to calculate some metrics on the NSQ candidates' research production, exclusively from open data, and to study if these metrics could give insights on the peer-review process.

The metrics we take into account are organized in two groups. Table 1 shows the first one. It contains some well-established measures to assess the publication success and academic activity of scholars, thus they do not require further explanation.

| *ID* | *Description* |
|---|---|
| **cand** | overall number of publications authored by the candidate found in the open sources of use |
| **books** | number of books authored by the candidate |
| **articles** | number of journal articles authored by the candidate |
| **other_pubbs** | number of other kinds of publications (e.g. proceedings articles and workshop papers) authored by the candidate |
| **co-au** | number of publications authored by both the candidate and at least one member of the commission |

Table 1. The basic metrics extracted from the list of publications of each candidate

The last metric in Table 1 (i.e. *co-au*) deserves some discussion. It is meant to investigate if the relation of co-authorship between the candidates and the commission members has played some role in the evaluation process. The idea of studying the co-authorship network between candidates and evaluators is not new. Bagues et al. (2019) and Martini et al. (2022) even applied it to the Italian NSQ studying how these connections relate to the potential candidates' decision to apply and their success in the NSQ. The authors extracted the co-authorship relations from all the publications submitted to the first round of the NSQ in 2012. Then, they measured the distance between the candidates and the commission members. They also considered two researchers as connected if they are affiliated to the same institution. The conclusion was that "the applicants tend to receive more favorable evaluations from connected evaluators". The authors performed a similar analysis to the Spanish Qualification (Zinovyeva & Bagues, 2015). In this work, they also took the connection <*PhD Student-Advisor*> into account. The findings are very similar to the Italian case and, even there, the "social ties" proved to have a great impact on the peer-review process.

Many other experiments have been performed with similar results in other contexts. Abramo et al. (Abramo et al., 2015a, 2015b) studied career advancements in the Italian academia. They considered the submissions in the Associate Professor competitions in 2008 and performed a statistical analysis to identify which factors have most determined the success of the candidates. They considered the research production (number of publications, citations, journals, etc.) but also the co-authorship and co-affiliation links between candidates and evaluators. The latter (same university) proved to be the most influential parameter. Several research groups also studied if co-authorship and co-affiliation connections increase the possibility of getting articles published in journals (Brogaard et al., 2014; Colussi, 2018; Dondio et al., 2019; Teplitskiy et al., 2018) or being appointed as journal editor (Miniaci & Pezzoni, 2020) or even receiving more research funds (Ebadi & Schiffauerova, 2015).
To the best of our knowledge, this is the first work that investigates citations as connectors between candidates and evaluators. Our other metrics, in fact, are calculated on the network of citations across the articles published by the candidates and the commission as reported in Table 2.

| *ID* | *Description* |
|---|---|
| **cand_comm** | number of citations going from a candidate's publication to a publication authored by at least one member of the commission |
| **comm_cand** | number of citations going from a publication authored by at least one member of the commission to a candidate's publication |
| **BC (bibliographic coupling)** | number of publications cited by both a publication authored by the candidate and a publication authored by at least one member of the commission |
| **CC (co-citation)** | number of publications citing both a publication authored by the candidate and a publication authored by at least one member of the commission |
| **cand_other** | number of other publications (i.e. which are not authored neither by the candidate nor by any member of the commission) cited by a publication authored by the candidate |
| **other_cand** | number of other publications (i.e. which are not authored neither by the candidate nor by any member of the commission) citing a publication authored by the candidate |

Table 2. The metrics extracted from the citation networks with a short description

Note that these indicators are not new, but they have never been studied from this perspective. Indeed, citation networks proved to be effective for topic identification and research trends analysis (Chang et al., 2015; Kleminski et al., 2020), as well as for studying interactions among research groups and institutions (Yan & Sugimoto, 2011), for evaluation purposes (Cai et al., 2019) and for visualizing research impact and production (Portenoy et al., 2017).

Here, the presence of citations from/to the publications of the candidates' and the commission indicates *proximity*. There are actually two forms of proximity embodied by the citations: cognitive and social. From the cognitive perspective, the fact that an author cites another one indicates that they share research interests and activities, they work in the same area and know their respective works. However, there is also a social dimension to take into account: candidates and commission members might work in the same department or university, might collaborate within the same national/international research project, might belong to the same research network, and their citations will probably be influenced by such connections.

This work does not make distinction between these forms of proximity, though some indicators that we took into account can be considered more on the cognitive side - for instance co-citation and bibliographic coupling - while others on the social side - for instance

cand_comm. Section “Looking at cognitive and social proximity in citations” contains some thoughts about such a distinction and possible directions for discerning between these forces within the citations.

Note also that it might well happen that the citation network of a candidate has a few connections with that of the commission, but the candidate equally gets a positive evaluation in the NSQ. In that case, there is no proximity considering only our metrics but other metrics might have influenced the evaluation.

There is another key point to stress on: our analysis is focused on open data. It might happen that these data are limited for some candidates and the list of their publications is underestimated and thus their citation network does have a very limited tie with that of the commission. In that case, the candidates have probably succeeded in the NSQ even if our data would suggest a different outcome.

To give readers a clearer idea of our approach, Figure 1 shows one example of the citation network we built for each candidate. A graph is created in which the nodes represent publications and the edges represent citation links. Note that we do not use arrows for the direction of the citations, just to not overload the picture, but we take it into account when computing the metrics. Different colors are used to indicate if a publication is by the candidate (blue) or the commission member (red) or co-authored by the candidate and at least one member of the commission (green).

The gray nodes indicate the publications of other authors - neither the candidate nor any member of the commission - that cite or are cited by a publication of the candidate or a member of the commission. These citations, in fact, contribute to the values of co-citations and bibliographic coupling. To make this evident, we colored these nodes depending on the CC and BC parameters: the more these values are high, the more the color is dark.
In the example in Figure 1, the candidate has a strong co-authorship connection with the commission members (half of her/his articles are co-authored with at least one member of the commission) but also a lot of candidate’s and commission members’ publications are closely-coupled by citation relationships, as confirmed by the dense network in the center of the diagram. Indeed, the candidate succeeded in the NSQ evaluation.

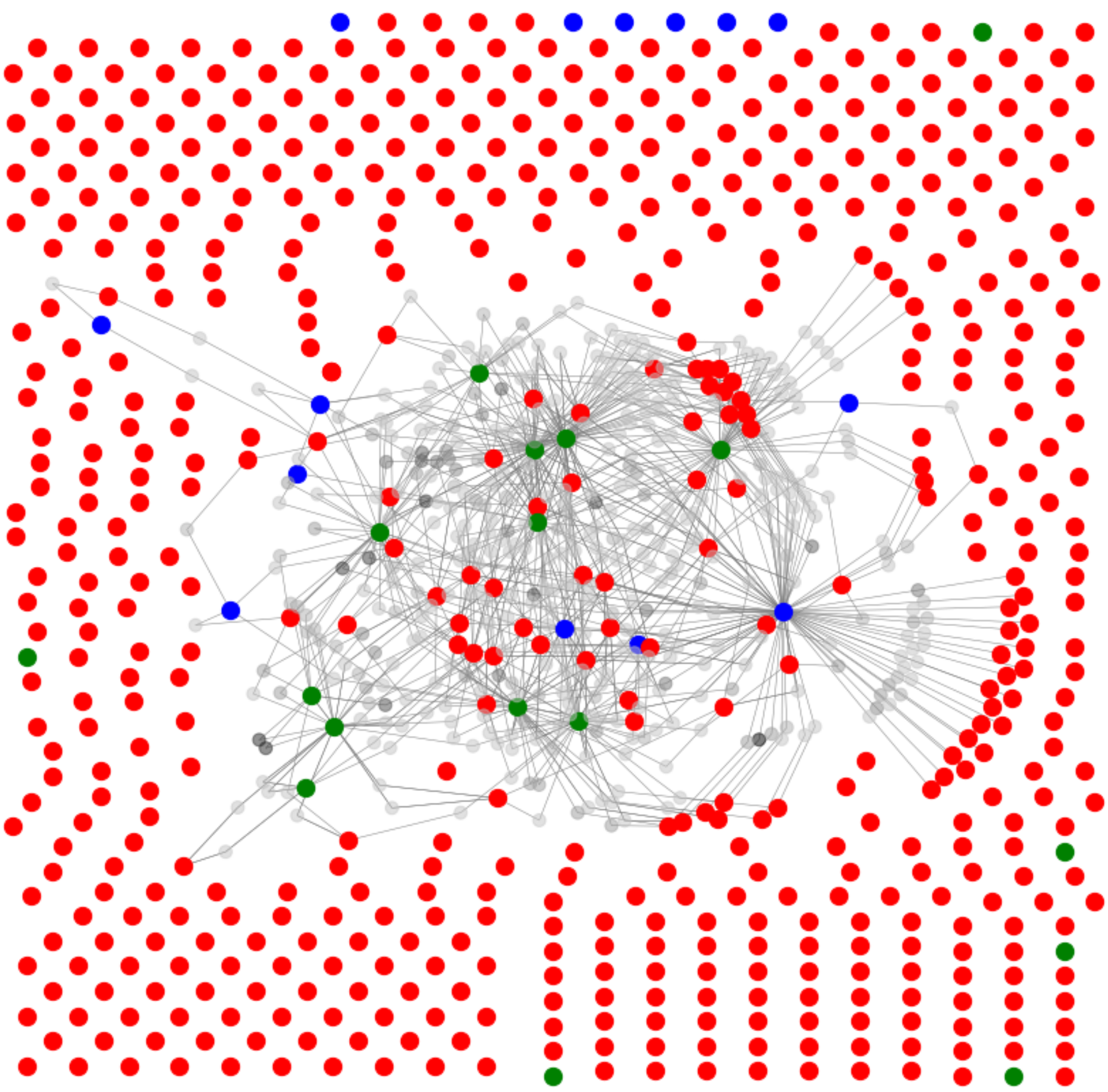

Figure 1: example of the kind of citation network we build for each candidate. The blue dots represent the candidate's publications, the red dots represent the commission's publications and the green dots represent the publications co-authored by the candidate and at least one member of the commission. The edges represent citation links. Gray nodes represent publications by other authors - neither the candidate nor a member of the commission - that connect a candidate's publication with a publication authored by at least one member of the commission via either bibliographic coupling or co-citation. The color of these nodes visually represents the number of publications that both the candidate's and the commission's publications either cite or are cited by: the darker the node, the higher the number of connecting publications.

Figure 2 shows the actual values of the metrics, in an alternative visualization loosely-based on Venn diagrams. The three colored sets represent the publications of the candidate (in light blue), of the commission (in pink) and of other authors (in light yellow, useful to calculate co-citation and bibliographic coupling). The set of publications co-authored by the candidate and at least one member of the commission are colored in violet, as the intersection of the two. The number of publications in each set is underlined while the numbers of outgoing/incoming citations among these sets are not; the grey rectangle is used to connect these values from and to each group. In this case, the candidate published 31 articles, including 17 co-authored with commission members, who cumulatively had 498 publications. The candidate's articles were also cited 8 times by the commission, while she/he cited 23 times articles authored by at least one member of the commission.

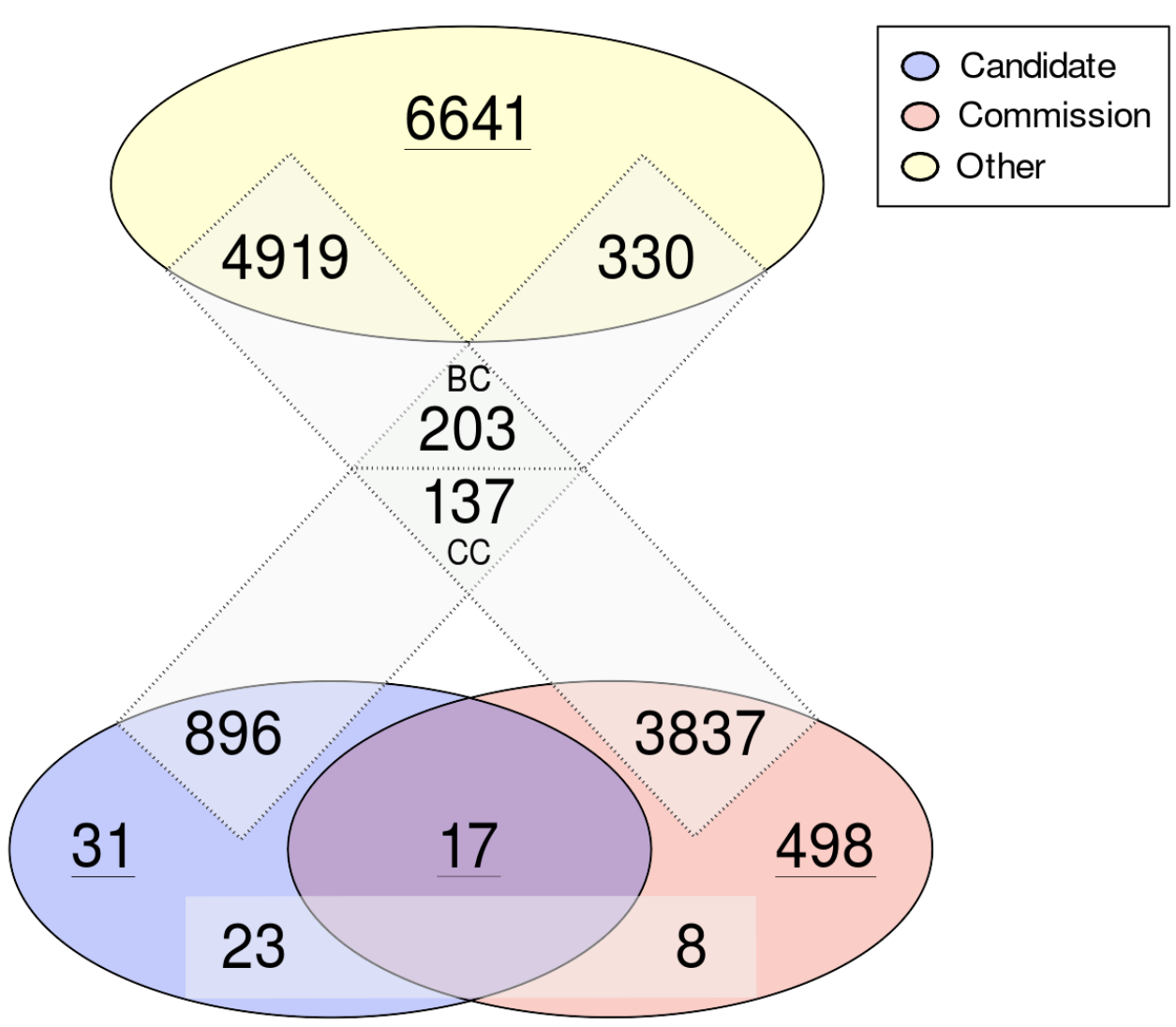

Figure 2: an alternative visualization for the co-authorship and citation network of Figure 1.

Figure 3 shows a different case, with both visualizations. The candidate has co-authored no articles with any of the members of commission, even if she/he has a good research record with 52 publications which received 666 citations. There is no citation from and to the candidate's articles and the articles authored by at least one member of the commission. Note that the data about the commission are different from the previous diagrams, since the two cases are from two different Recruitment Fields. In this case the candidate failed the NSQ.

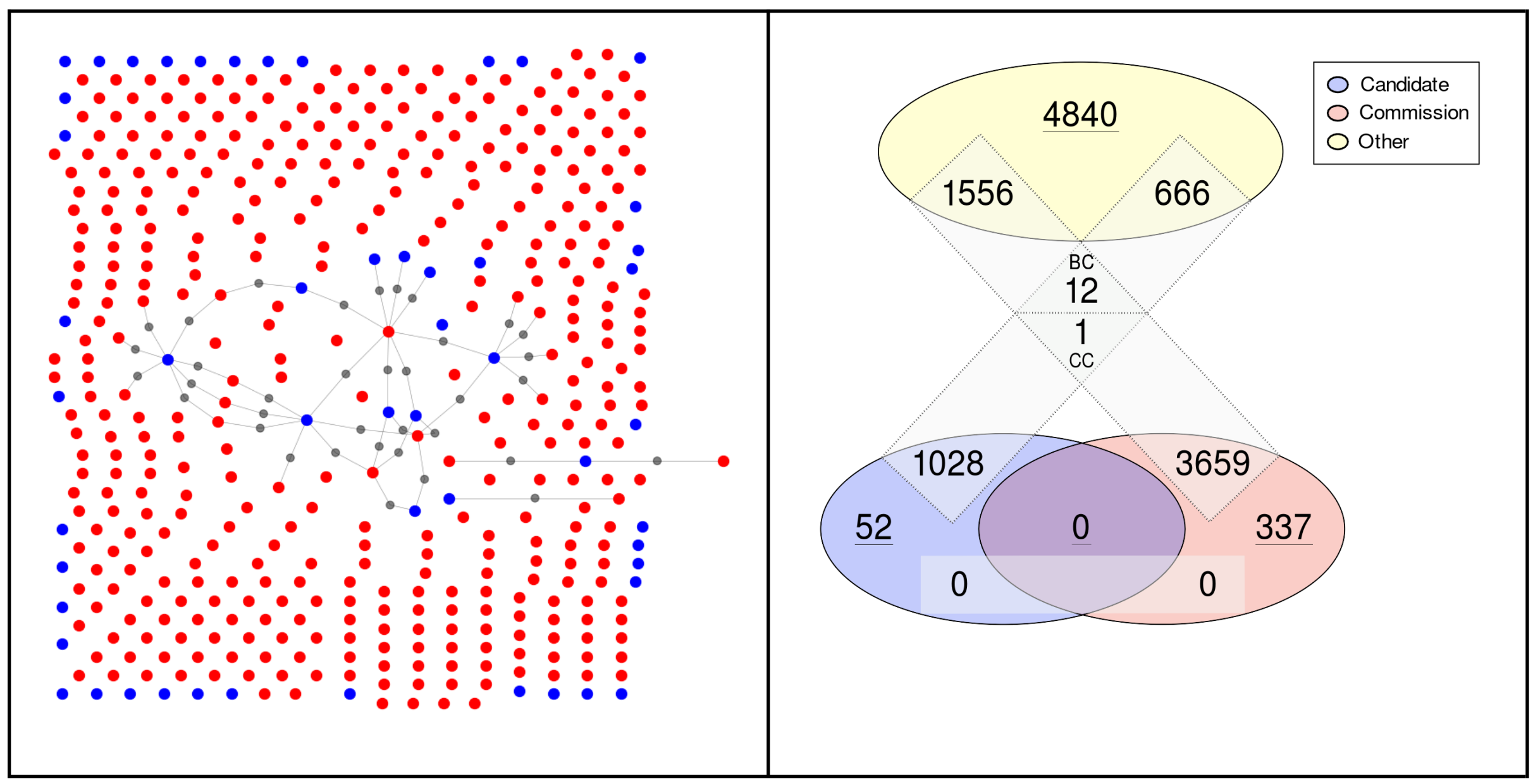

Figure 3: The case of a candidate with no direct connection with the commission members and a good research record.

We built all the citation networks and computed our bibliographic metrics using exclusively open data, in order to test whether quantitative citation-based indicators can deepen our understanding of peer-reviewed NSQ evaluation. The following subsections describe in detail the data collection process and the sources we used.

**Selection of Recruiting Fields and Data**

For the purposes of this study, we considered the bibliographic data of the candidates and commissions that applied to the Recruiting Fields 10/G1, *Historical and General Linguistics*, and 13/D4, *Mathematical Methods of Economics, Finance and Actuarial Sciences* in the 2016, 2017, and 2018 sessions of the NSQ. In order to choose which 2 RFs we should focus on, we collected all the bibliographic information related to candidates to the role of AP of the 2016 NSQ session and compared the coverage of all the RFs in Microsoft Academic Graph (referred to as MAG). Since the majority of NDs are humanistic disciplines, we decided to base this comparison on the bibliographic data found in MAG. Indeed, previous studies have demonstrated that it has a better coverage of such disciplines (Martín-Martín et al., 2020).

For each candidate, we calculated the percentage of publications found in MAG by dividing the number of all publications in MAG by the number of publications listed in his/her CV submitted to the NSQ. Then, for each RF, we calculated the median percentage. In this comparison, coverage percentages can go well above 100%, as we also considered any additional publication by the same author that we retrieved from MAG but was not included in the candidate's CV. Candidates can decide not to list all of their academic production on their CV, but only their most recent or relevant publications. However, these remaining ones can be found in MAG and added to the count. We decided to also consider these additional publications since all of a candidate's bibliographic and citation data is useful for the calculation of our metrics. Among the top 15 RFs in terms of median percentage of publications found in MAG, 10/G1 and 13/D4 were the best covered two RFs that do not belong to the same SA. 13/D4 being the first RF in terms of coverage with 269% publications found in MAG and 10/G1 being the 15th with 156%. All the remaining RFs from the 2nd to the 14th position belonged to SA 13.

**Sources**

In October 2020, we collected open citation data making use of four sources. The first dataset we used is Microsoft Academic Graph (referred to as MAG here), which results from the efforts of the Microsoft Academic Search (MAS) project. This dataset is updated biweekly, is distributed under an open data license for research and commercial applications and is accessible using the Project Academic Knowledge API (https://docs.microsoft.com/en-us/academic-services/project-academic-knowledge/introduction)(Wang et al., 2020).

The second dataset we used is the OpenAIRE Graph (referred to as OA here), which includes information about objects of the scholarly communication life-cycle (publications, research data, research software, projects, organizations, etc.) and semantic links among them. It is created bi-monthly, and is freely accessible for scholarly communication and research analytics using the OpenAIRE Graph Access API (http://api.openaire.eu/)(Rettberg & Schmidt, 2012).

The third is Crossref (referred to as CR here), which was born as a nonprofit membership association among publishers to promote collaboration to speed research and innovation. The dataset is fully curated and governed by the members as they autonomously provide their publications' DOIs, webpages, metadata, and update the information whenever it is necessary. The metadata are available through a number of APIs, including REST API (https://github.com/CrossRef/rest-api-doc) and OAI-PMH (Hendricks et al., 2020).

Lastly, we also relied on COCI, the OpenCitations Index of Crossref Open DOI-to-DOI Citations (Heibi et al., 2019), made available by OpenCitations (Peroni & Shotton, 2020). COCI only stores the citation links between the citing and cited bibliographic entities identified by their DOIs and enables their metadata to be retrieved from other sources using the corresponding REST API (https://opencitations.net/index/coci/api/v1).

**Collection process**

In order to calculate citation-based metrics that quantified the overlap between the candidate and the commission's citation networks, we needed to collect the bibliographic and citation data of both the candidates and the commissions. To achieve this result, we used the four step procedure shown in Figure 4 and described below.

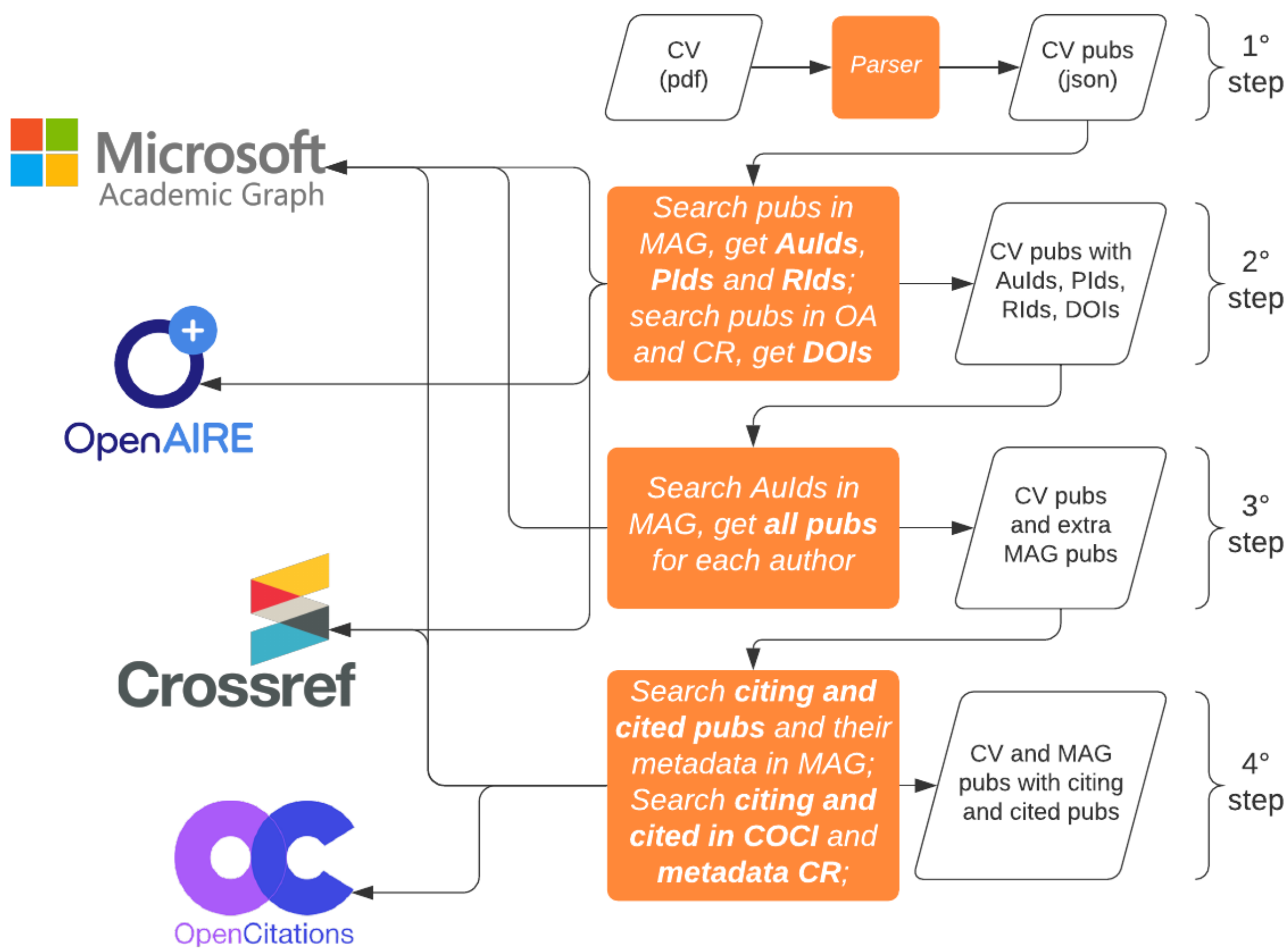

Figure 4: data collection process stages with resources and outcomes. MAG, OA, CR, and COCI stand for Microsoft Academic Graph, OpenAIRE, Crossref and OpenCitations Index of Crossref Open DOI-to-DOI Citations. AuId, PId, and RId stand for Author Id, Paper Id, and Reference Id respectively. The first identifies the author, the second identifies the entity and the third is a list of the Paper Ids of all cited entities.

First, we extracted the publications' metadata. For candidates, we extracted the metadata of each publication listed in the candidates' resumes. In particular, the metadata contained in the resumes available in PDF format were stored into JSON files. In the resumes, the candidate's publications are listed under two sections: "publications" and "publications for indicators". The first are considered in the peer-review phase. Whereas, the second are utilized to compute the metrics used in the preliminary phase of the NSQ. Since there is often a partial overlap between these two lists, publications were disambiguated based on DOI, when present, or title and year correspondence. To ensure matching, space, accents, and punctuation were removed from the publications' titles. For the commissions, we manually compiled a list of publications and searched for their metadata for each member. Since all commissions' members are professors, we searched for their IRIS web pages[4] in the browser. For each member it was possible to find their IRIS publications' page inside their university's IRIS website. The interface allows for the download of the publication's metadata in Excel format. These files were then combined into a single CSV which stored all the publication data of the commissions.

Secondly, we searched for each candidate's Author Ids in MAG and for each article's DOI in OA and CR. In MAG, each author may be assigned with multiple Author Ids due to author disambiguation issues. Therefore, we queried MAG for each article in order to retrieve as many Author Ids as possible. Each original CV article was searched for by title and year or, when present, by DOI, using the *Evaluate* method of the MAG REST API. Academic entities returned by DOI query were considered valid when there was exact lowercase correspondence between the original and found DOI – since DOIs are case insensitive identifiers. In the case of title and year queries, returned academic entities were matched to the original articles based on:

- title correspondence (after having removed punctuation, accents and spaces from the titles);
- publication date (a two year margin is permitted);
- author's surname and author's name (only in case two or more authors shared the same surname).

Once the correspondence between the returned and the original academic entity had been validated, the following attributes were retrieved from MAG: Author Id, Paper Id, and Reference Ids. The first identifies the author, the second identifies the entity and the third is a list of the Paper Ids of all cited entities. The remaining publications that were not found in MAG and were not associated with a DOI were searched for in OA by title's keywords, author's surname and publication date. Keywords were chosen by selecting the first six words in the title that were not stopwords. Since OA strictly returns publications that perfectly match with the attributes specified in the query, the first returned entity was considered valid and the corresponding DOI was retrieved. Lastly, publications that had not been found in OA and did not have a DOI were searched for in CR by title's keywords and surname. We considered only the first four results and we calculated three similarity scores for each. Score

[4] IRIS is an IT solution that supports Italian universities in collecting, storing and managing research data.

A was calculated based on how close the publication date was to the original one: 3 points were given if the new publication had the same date of the original; 2 points if the new publication's date was one year apart from the original's; 1 point if the new publication's date was two years apart from the original's. Score B was based on the surname and name correspondence: 2 points were given if the new publication's author had the same name and surname of the original's; 1 point was given if the new publication's author had the same name and initial of the original's. Score C was the Levenshtein distance between returned title and original title. The results were then ranked by C score. If the result with the highest C score has a C score greater than 0.8 and has A and B scores equal to or greater than 1, then the entity is considered valid and the corresponding DOI is retrieved. This matching solution was inspired by Visser, van Eck and Waltman's own matching process (2020).

Thirdly, MAG was queried by each Author Id and the returned entities that were not already contained in the CV were added. Each returned entity was compared to the CV publications by Paper Id, DOI or title and year. If the new publication matched the present one, the missing information would be added and the new publication would be discarded. In case the new entity did not match any previous publication, it would be included in the JSON file under a different category than the publications originally present in the CV. It would be categorized as an extra publication found in MAG.

In the fourth step of this procedure, the metadata of the citing and cited publications were retrieved. For those publications having at least one Reference Id, the metadata of the entities cited by that publication was collected by querying MAG by Paper Id using each Reference Id. Viceversa, citing entities metadata was returned if querying MAG by Reference Id using the publication Paper Id. To ensure the best coverage of citation data, all the publications associated with a DOI were searched in COCI to collect the DOIs of the citing and cited publications. These were then queried in CR to retrieve their metadata. Given that the same citing or cited publication could be retrieved from both MAG and COCI, the citing and cited articles were disambiguated by DOI to prevent repetitions.

As summarised in Table 3, 500 unique NSQ applications spread over 5 terms from 433 unique candidates were considered. The number of unique applications is greater than the number of unique candidates due to the fact that candidates can re-apply to the NSQ if rejected. Overall we obtained 15,753 publications from MAG, OA, CR and COCI. In particular, we found 11,617 of the original 15,330 publications listed in the candidates CVs.

| **500** unique applications |
| --- |
| **433** unique candidates |
| **10** unique commission experts |
| **15,330** candidate publications listed in their CVs |
| **11,617** candidate CV publications found in open data sets |
| **15,753** candidate publications found in open data publications |
| **865** commission publications found in open datasets |

Table 3. Dataset specifics

Finally, applications were divided into three sections, A, B or C, depending on how many of the original CV publications were retrieved from the mentioned open access datasets. Applications were placed in section A when more than 15 of the publications originally listed in the CV were retrieved, or, otherwise, 70% of the CV publications were retrieved. Applications were placed in section B when less than 70% of their CV publications were found but additional publications not listed in the CV were extracted from MAG to reach a total number of retrieved publications comparable to the original one. Applications were placed in section C otherwise. This categorization gives us the ability to decide whether to include applications that are not greatly covered in our datasets of use in our experiment, described in the following section.

**Coverage in open access datasets: MAG, OpenAIRE and Crossref**

Before running the machine learning classifiers, we examined the coverage of the publications of the candidates for the RFs 10/G1 and 13/D4 in the three datasets we took into account: MAG, OpenAIRE and Crossref . This helped us to better understand our results and to identify some weaknesses of these datasets.

To do so, we queried the datasets for each unique publication listed in the CV of each candidate. We then calculated the percentage of unique CV publications found in each dataset over the total number of unique publications in the CV. As shown in Figure 5, among the three datasets, the one with the best coverage of both sections and fields is MAG. However, combining the three datasets yields the best result, which means that some bibliographic entities included in other datasets were not included in MAG.

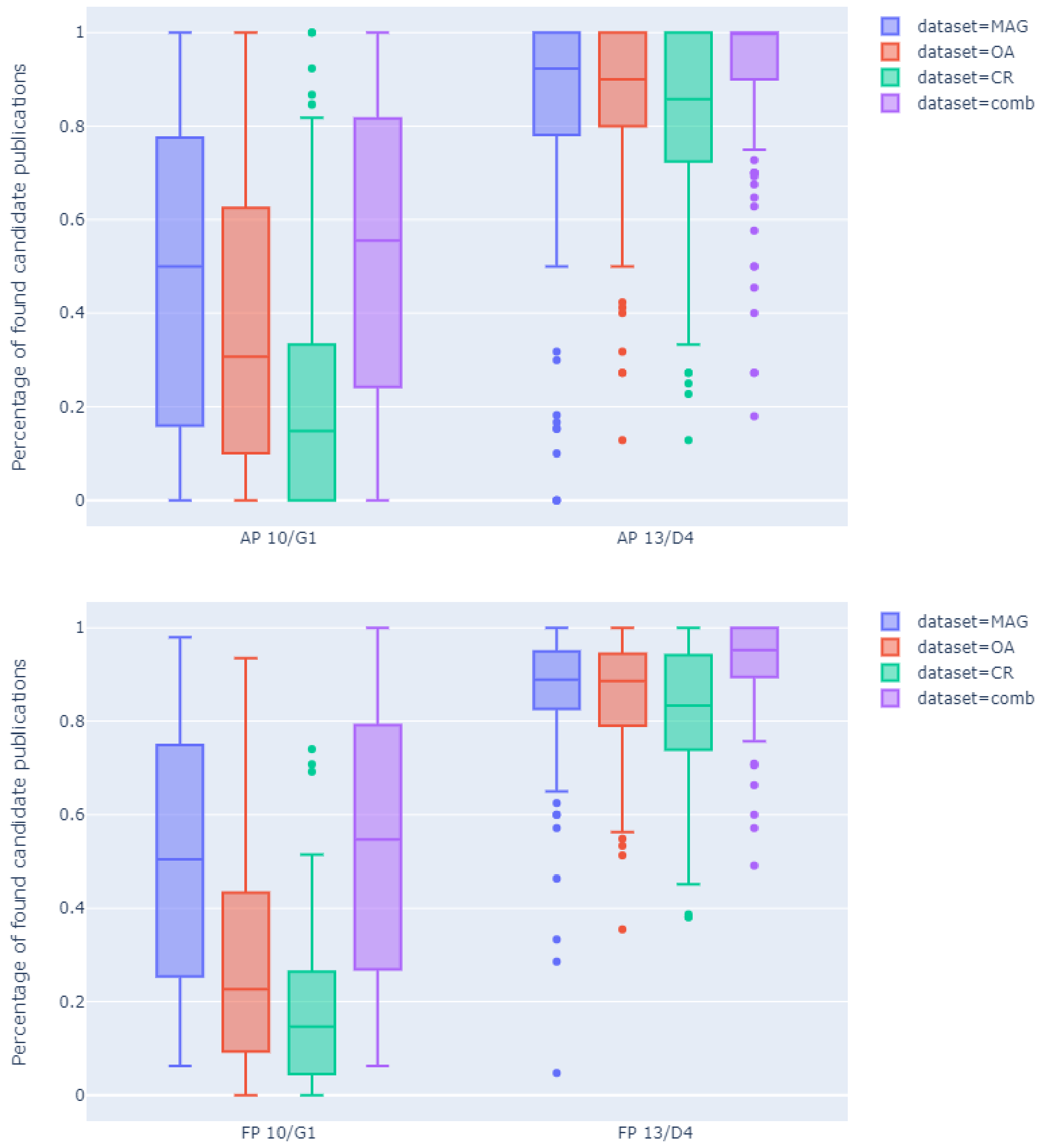

Figure 5: boxplots representing the coverage of each dataset for each role and field. FP and AP stand for Full Professor and Associate Professor respectively.

In the case of the RF 13/D4, MAG's coverage is slightly higher or equal to that of OA and CR. Whereas for 10/G1, MAG's coverage almost doubles that of OA and more than doubles that of CR. Despite both being categorized as a ND, 10/G1 is more traditionally humanistic than 13/D4: the first is interested in linguistics, while the second in mathematical models. MAG's better coverage of publications and citations in the Humanities is also confirmed by the literature on the topic (Martín-Martín et al., 2020).

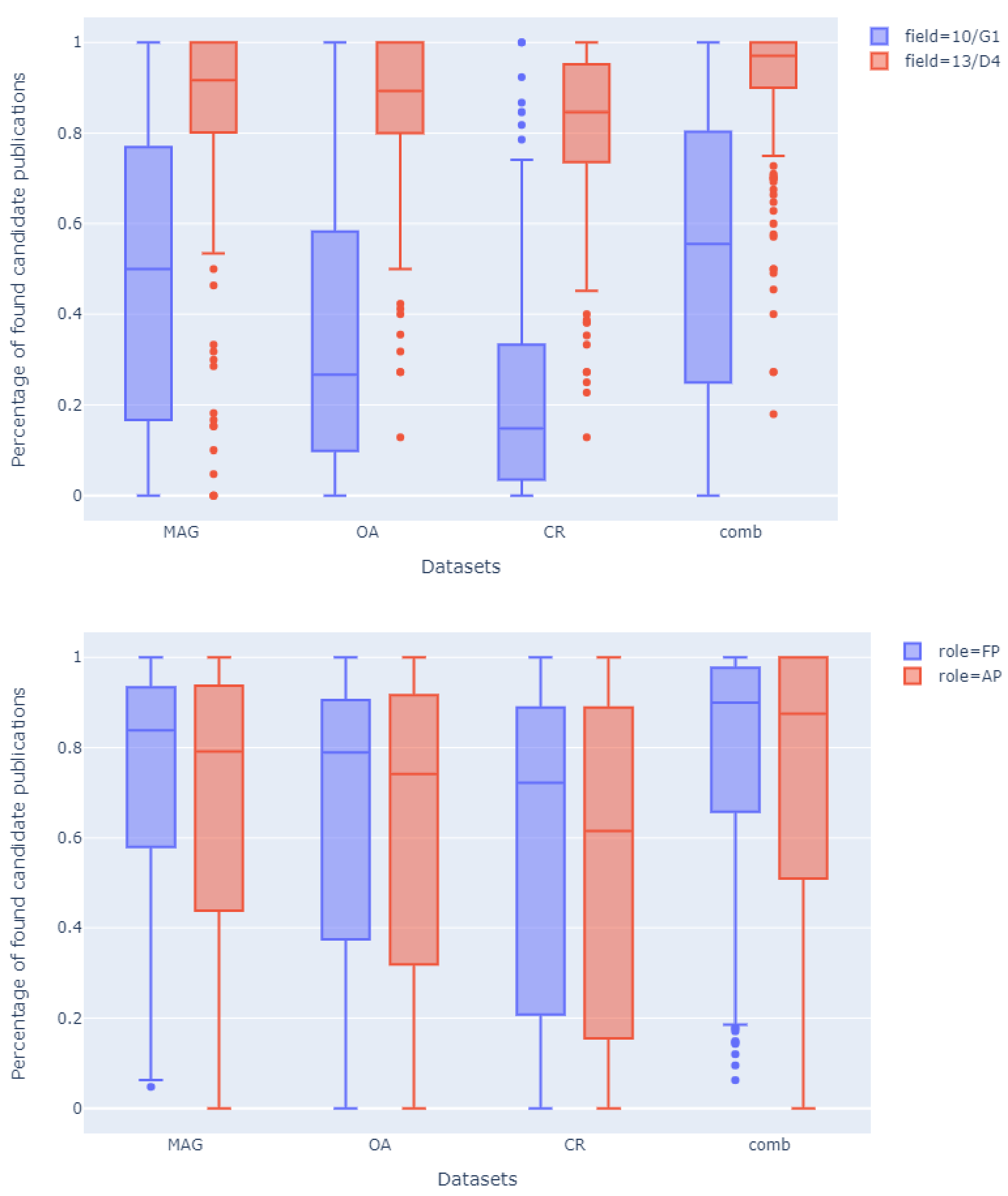

Figure 6: boxplots representing the coverage of each dataset for role and each field. FP and AP stand for Full Professor and Associate Professor respectively.

Interestingly, the datasets' coverage varies more greatly across the two RFs than across the two academic roles (i.e. FP and AP), as shown in Figure 6. Hence, coverage of a candidate's publications is not affected by what role the candidate is applying for, but rather by their field of study. This indicates that the datasets cover scientific fields better than humanistic ones.

It is worth pointing out that these percentages are evidently lower than those presented in Section "Selection of Recruiting Fields and Data" because any additional publication to those in the CV found in MAG was not taken into consideration in this analysis, in order to ensure a fair comparison between the datasets.

**Machine Learning Classifiers**

The aim of the analysis we present is to investigate whether citation-centric metrics can provide insights on the human evaluation in the NSQ of NDs. To this end, we designed an experiment based on computational methods and machine learning techniques. In particular, we employ two different machine learning classifiers: SVM and Decision Trees. These classifiers have been recently used in many bibliometric studies, demonstrating their effectiveness in different contexts. For instance, SVM has been used to predict academic career outcomes (Tregellas et al., 2018) and the results of evaluation procedures of academics (Nuzzolese et al., 2019; Poggi et al., 2019) based on bibliometrics indicators, to perform citation analyses for identifying instrumental citations (Fu et al., 2013), to recognize in advance articles that will become relevant and breakthrough (Savov et al., 2020). Also Decision Trees have been extensively used, for example to measure scientific knowledge flows using citation context analysis (Hassan et al., 2018), to estimate and predict article citation count based on bibliometric data (Fu & Aliferis, 2010) and text analysis techniques (Ibáñez et al., 2009).

Our experiment is based on a comparison of the behaviors of the metrics introduced in Table 1 and Table 2, and those provided by ANVUR, described in Section "The National Scientific Qualification: how it works". These metrics are used as features of classification algorithms for automatically discerning between candidates who passed the NSQ and who did not. This is a binary classification problem since we have two mutually exclusive classes (i.e. one composed of candidates who passed the NSQ, and the other of candidates who failed) and, for each individual in the population, we attempt to predict which class he/she belongs to.

We base our analysis on a variable ranking method from feature selection theory (Guyon & Elisseeff, 2003) to determine the metrics that are significant for the current classification problem, and hence are useful to discriminate between candidates passing the NSQ or not, and to identify the metrics that are either redundant or irrelevant – and that can, thus, be removed without incurring much loss of information (Bermingham et al., 2015).
The first step of the experiment consists in splitting the data about the NSQ candidates into a *training set*, composed of the information about 283 candidates from the first four NSQ terms (56.6% of the total), and a *test set,* composed of the information on the remaining 217 candidates from the last term (43.4% of the total). In this way we use around half of the data to train the algorithms, which are tested on the remaining data.

For each academic recruitment field (i.e. 10/G1 and 13/D4) and academic role (i.e. FP and AP), we have two nominal classes to use for the two classification algorithms (SVM and Decision Trees) that come from the NSQ procedure, i.e. "Passed" or "Failed". Table 4 reports the details about the dataset used for experimenting with the classifiers.

| | FP | | | AP | | |
|---|---|---|---|---|---|---|
| | Passed | Failed | Total | Passed | Failed | Total |
| 10/G1 (training) | 22 | 13 | 35 | 29 | 35 | 64 |
| 10/G1 (test) | 13 | 22 | 35 | 16 | 23 | 39 |
| 13/D4 (training) | 35 | 41 | 76 | 22 | 17 | 39 |
| 13/D4 (test) | 53 | 55 | 108 | 24 | 36 | 60 |

Table 4. Dataset for the classification experiment

We test each possible subset of metrics and find those which minimizes the error rate and lead to good predictions. We start by computing all the possible combinations of metrics that can be obtained considering the eleven metrics based on open datasets and the three indicators provided by ANVUR. The number of combinations can be computed by the following expression:

$$\sum_{k=1}^{14} C_{14,k} = \sum_{k=1}^{14} \frac{n!}{k!\,(n-k)!} = 16{,}383$$

For each of the two academic recruitment fields (i.e. 10/G1 and 13/D4), two academic roles (i.e. FP and AP), and three input data coverages (i.e. A, AB and ABC), all the combinations of metrics have been used as features of the classification algorithms for automatically predicting the results of the NSQ. We consider Decision Trees and SVM with regularization parameter C[5] set to 1.0, 0.5, 0.1 and 0.02, resulting in five classification algorithm configurations. The total number of classifiers computed is 982,980, as summarized in Table 5.

| variable description | number of instances/counter |
| --- | --- |
| combinations of metrics | 16,383 |
| recruitment fields (i.e. 10/G1 and 13/D4) | 2 |
| academic roles (i.e. FP and AP) | 2 |
| input data coverages (i.e. A, AB and ABC) | 3 |
| classification algorithm configurations (i.e. Decision Trees and SVM with regularization parameter C set to 1.0, 0.5, 0.1 and 0.02) | 5 |
| **Total number of classifiers to compute** | **982,980** |

Table 5. Total number of classifiers to compute for the experiment.

All these classifiers are trained on the data about candidates to the first four terms of the NSQ, and are tested on the candidates of the last term. An oversampling technique with stratification has been applied to the minority class for managing imbalanced classes[6], i.e. in cases where the distribution of examples across the known classes is not equal (Krawczyk, 2016). In particular, we used the approach described in (Japkowicz, 2000) that prescribes to resample the smaller class (by creating synthetic instances chosen at random from the minority class) until it consisted of as many samples as the majority class. The Python code developed for this experiment is based on the Pandas[7] and Scikit-learn[8] libraries, and is available at https://github.com/sosgang/bond.

[5] SVMs work by finding data points of different classes (i.e. the support-vectors) and drawing boundaries between them (i.e. the hyperplanes). C behaves as a regularization parameter that trades off the correct classification rate of training examples against the maximization of the decision function's margin. For larger values of C, a smaller margin will be accepted if the decision function is better at classifying all training points correctly. A lower C will encourage a larger margin for future data at the cost of training accuracy. Since there is no general rule in choosing the C parameter as this depends on the dataset in use, we decide to test different values ranging from large to smaller ones for C in our experiment. An interesting discussion with examples on the topic is available at https://stats.stackexchange.com/questions/31066/what-is-the-influence-of-c-in-svms-with-linear-kernel#answer-159051.

[6] Machine learning algorithms are designed to improve accuracy by reducing the overall error. This can be a problem when classes are strongly imbalanced, since in these cases machine learning classifiers tend to be very biased and inaccurate towards the minority class. For instance, if we had two classes A and B and a population

Of all the computed classifiers, we consider only those with good discrimination abilities (i.e. those whose weighted average F1-score is at least 0.7, and which therefore lead to good classification performances). Since the classifiers were calculated using all possible combinations of metrics, each metric was used as a feature by half of the classifiers. We then count how many times each of the fourteen metrics has been used as a feature by the classifiers with good discrimination abilities, and identify as significant for the current classification problem those that have been used by more than 50% of these classifiers. We also identify the metrics that are not relevant or redundant as those that are used as features by a low fraction (i.e less than 35%) of the classifiers with high classification performances.
The results of this experiment allow us to understand what metrics provide relevant information that can be used by machine learning algorithms to replicate the decisions of the NSQ commissions. The hypothesis here is that if a subset of metrics is used by most of the classifiers with good performances, it signals that such metrics are important proxies of the human evaluation process and outcome.

## Results

The results of the 982,980 computed classifiers described in the previous section, in terms of precision, recall and F1-score are available in (Bologna et al., 2021) and in the GitHub repository https://github.com/sosgang/bond. Of all the computed classifiers, 4,217 have good discrimination abilities, with F1-scores spanning from 0.700 (i.e. the threshold) to 0.807 (i.e. the F1-score of the best classifier). The number of times each of the fourteen metrics has been used as a feature by these classifiers is reported in Table 6, together with their relative

---

of 100 individuals of which 95 belong to class A and the remaining five to class B, a silly classifier predicting always class A would have an overall accuracy of 95%, but an error rate for class B of 100%. This behavior is an issue in many scenarios, for instance anomaly detection, identification of fraudulent transactions in banks, identification of rare diseases, etc., and many techniques have been developed for solving or mitigating this problem.

[7] https://pandas.pydata.org/

[8] https://scikit-learn.org/

percentages. Table 6 also reports the results for each recruitment field and academic role, allowing us to perform a more specific analysis and get insights into the different relevance of metrics in the two disciplines and the two academic roles. It is however important to underline that in 13/D4 the number of classifiers with good discrimination abilities is very low compared the number of classifiers in 10/G1 (i.e. 117 vs 4100).

Among the citation-based metrics, *cand_comm* and *comm_cand* have been used in around 50% of the classifiers for both fields and academic roles ( 0.04% in 10/G1 FP, 50.26% in 10/G1 AP, 52.88% in 13/D4 FP and 51.06% in 13/D4 AP for both metrics). Whereas, *BC* has been used in more than half of the classifiers for 10/G1 FP (71.09%), but in very few ones in 10/G1 AP and 13/D4 AP, and in no classifiers in 13/D4 FP. *CC* has a very high percentage of use in both disciplines (99.65% in 10/G1 and 85.11% in 13/D4) at the AP level. Whereas, at the FP level, it has a percentage above average (i.e. 57.58%) in 10/G1 and a very low percentage (i.e. 28.57) in 13/D4. *cand_other* has low percentages in both fields and both roles, and *other_cand* is used in more than 50% of the classifiers only in 13/D4 AP (68.09%).

Of the non-citation-based metrics, *cand* is used in 66.61% and 71.43% of the classifiers in both fields for the FP role, whereas it is used in less than 50% of the classifiers for the AP role. Interestingly, *co-au* have above-average percentages in 10/G1 AP and 13/D4 FP (50.26% and 60% respectively), but low percentages in 10/G1 FP and 13/D4 AP. *books* has above average percentages in both fields and both roles, but 10/G1 FP (51.81% 10/G1 AP, 55.71% 13/D4 FP, 82.98% 13/D4 AP). *articles* has low percentages in both fields and both roles, whereas *other_pubbs* reaches an above average percentage only in 13/D4 FP (75.71%).

All ANVUR metrics are used in less than half of the classifiers for both roles in 10/G1. However, *ND_M1* and *ND_M2* have higher-than average percentages for 13/D4 AP (51.06% and 85.11 respectively), and ND_M3 is in half of the classifiers for 13/D4 FP (50%).

| | | | METRICS BASED ON OPEN DATASETS | | | | | | | | | | | ANVUR METRICS | | |
|---|---|---|---|---|---|---|---|---|---|---|---|---|---|---|---|---|
| | | | CITATION-BASED METRICS | | | | | | NON-CITATION-BASED METRICS | | | | | NON-CITATION-BASED METRICS | | |
| | | | cand_ comm | comm_ cand | BC | CC | cand_ other | other_ cand | cand | co-au | books | articles | other_ pubbs | ND_M1 | ND_M2 | ND_M3 |
| **10/G1 FP** | **1207 classi fiers** | feature count | 604 | 604 | 858 | 695 | 384 | 452 | 804 | 512 | 536 | 136 | 600 | 188 | 72 | 487 |
| | | feature percentage | **50.04%** | **50.04%** | **71.09%** | **57.58%** | 31.81% | 37.45% | **66.61%** | 42.42% | 44.41% | 11.27% | 49.71% | 15.58% | 5.97% | 40.35% |
| **10/G1 AP** | **2893 classi fiers** | feature count | 1454 | 1454 | 1080 | 2883 | 883 | 704 | 1423 | 1454 | 1499 | 1147 | 523 | 758 | 516 | 962 |
| | | feature percentage | **50.26%** | **50.26%** | 37.33% | **99.65%** | 30.52% | 24.33% | 49.19% | **50.26%** | **51.81%** | 39.65% | 18.08% | 26.20% | 17.84% | 33.25% |
| **13/D4 FP** | **70 classi fiers** | feature count | 37 | 37 | 0 | 20 | 32 | 5 | 50 | 42 | 39 | 0 | 53 | 20 | 0 | 35 |
| | | feature percentage | **52.86%** | **52.86%** | 0.00% | 28.57% | 45.71% | 7.14% | **71.43%** | **60.00%** | **55.71%** | 0.00% | **75.71%** | 28.57% | 0.00% | **50.00%** |
| **13/D4 AP** | **47 classi fiers** | feature count | 24 | 24 | 7 | 40 | 4 | 32 | 11 | 7 | 39 | 19 | 19 | 24 | 40 | 18 |
| | | feature percentage | **51.06%** | **51.06%** | 14.89% | **85.11%** | 8.51% | **68.09%** | 23.40% | 14.89% | **82.98%** | 40.43% | 40.43% | **51.06%** | **85.11%** | 38.30% |

**Table 6**. The results of the experiment for each recruitment field and academic role. The metrics with high percentages of use are highlighted in green.

## Discussion and conclusions

In this work, we explored whether open citation-based metrics relate to the qualitative peer-review of NDs. Specifically, we grounded our analysis on the data of the candidates and commissions that took part in the 2016, 2017, and 2018 terms of the NSQ for the disciplines *Historical and General Linguistics* and *Mathematical Methods of Economics, Finance and Actuarial Sciences*, having Recruiting Fields 10/G1 and 13/D4. We collected the bibliographic metadata and citation data from MAG, OA, Crossref and OpenCitations. We devised and reused a set of bibliographic and citation-based metrics using a top-down process and combining popular metrics (e.g. bibliographic coupling) to others proposed for the purpose of this study to describe the relationship between the candidates and the commision of the NSQ.

The results of the experiment presented in Section "Results", based on computational methods and machine learning techniques, provides insights into which metrics are relevant in the human evaluation of different disciplines and academic levels. Our hypothesis is that metrics used by most of the classifiers with good performances to predict the outcome of the NSQ, such metrics are important proxies of the human evaluation process.

First of all, we remark that citation-based metrics register high percentages of use in classifiers with good discrimination abilities in both disciplines and academic roles, i.e. *cand_comm*, *comm_cand*, *BC* and *CC*. The relevance of citation-based metrics in both disciplines is the first element supporting our hypothesis.

#### Analysis of Citation-Based-Metrics

Regarding our citation-based metrics, *cand_comm* and *comm_cand* show an average percentage of use in all disciplines and levels. Since these metrics describe the strength of the citational relationship between the candidate and the commission, their high percentage of use in the classifiers appears to indicate that this relationship does play a role in the peer-review phase of the NSQ. Therefore, it suggests a positive answer to the second part of our research question. Additionally, both *CC* heightened relevance at the AP level for both disciplines and *other_cand*'s high percentage in 13/D4 AP suggest that being recognized as a member of the same scientific community of the commission, or as a trust-worthy researcher by a third party weighs heavily in the evaluation of junior scholars. Indeed, *CC* is the number of publications citing both the candidate and the commission, and *other_cand* is the number of citations from others to publications of the candidate. Furthermore, in 10/G1 FP, *BC* has a high percentage of use, and *CC* has an above-average percentage. This suggests that the evaluation of candidates applying to this discipline and academic role is based on the strength of their relationship with the commission and their connections within their scientific community. In order to further investigate CC relevance for the academic role AP, we analyzed the structures of the computed Decision Trees.

Decision Trees are flowchart-like structures composed of nodes (represented by rectangles) and branches (represented as arrows connecting nodes) showing the flow from question to answer. Each internal node represents a test on a feature (e.g. whether a coin flip comes up heads or tails), each branch represents the outcome of the test, and each leaf node represents a class label (decision taken after evaluating all tests). For instance, Figure 7 is the Decision Tree generated using the four features *CC, books* and *nd_m1*, and applied to the test set data having coverage A or B. Internal nodes are depicted as gray rectangles, and report the tests and the number of individuals in the test set belonging to each class (i.e. "pass" for candidates who passed the NSQ and "fail" for those who did not) between square brackets. Leaf nodes are depicted as light blue rectangles, and report the output class labels, the actual number of individuals in the test set belonging to each class and the node accuracy.

By analyzing the Decision Tree we observe that the first test splits individuals having *CC* greater than one (right branch) from those having *CC* lower or equal to one (left branch). The former flow into a leaf node (i.e no other condition is tested), and are labeled as candidates who did not pass the NSQ resulting in an accuracy score of 0.75. The latter flow in the left branch, requiring a series of tests on the metrics *books, nd_m1* and *CC* to discern different cases, and leading to high accuracy values in all of them.

The Decision Trees generated for this sector share similar structures, which emphasizes strong significance for the CC metric. In fact, with just one test about the metric *CC* it is possible to correctly label with good accuracy almost half of the individuals in the test set.

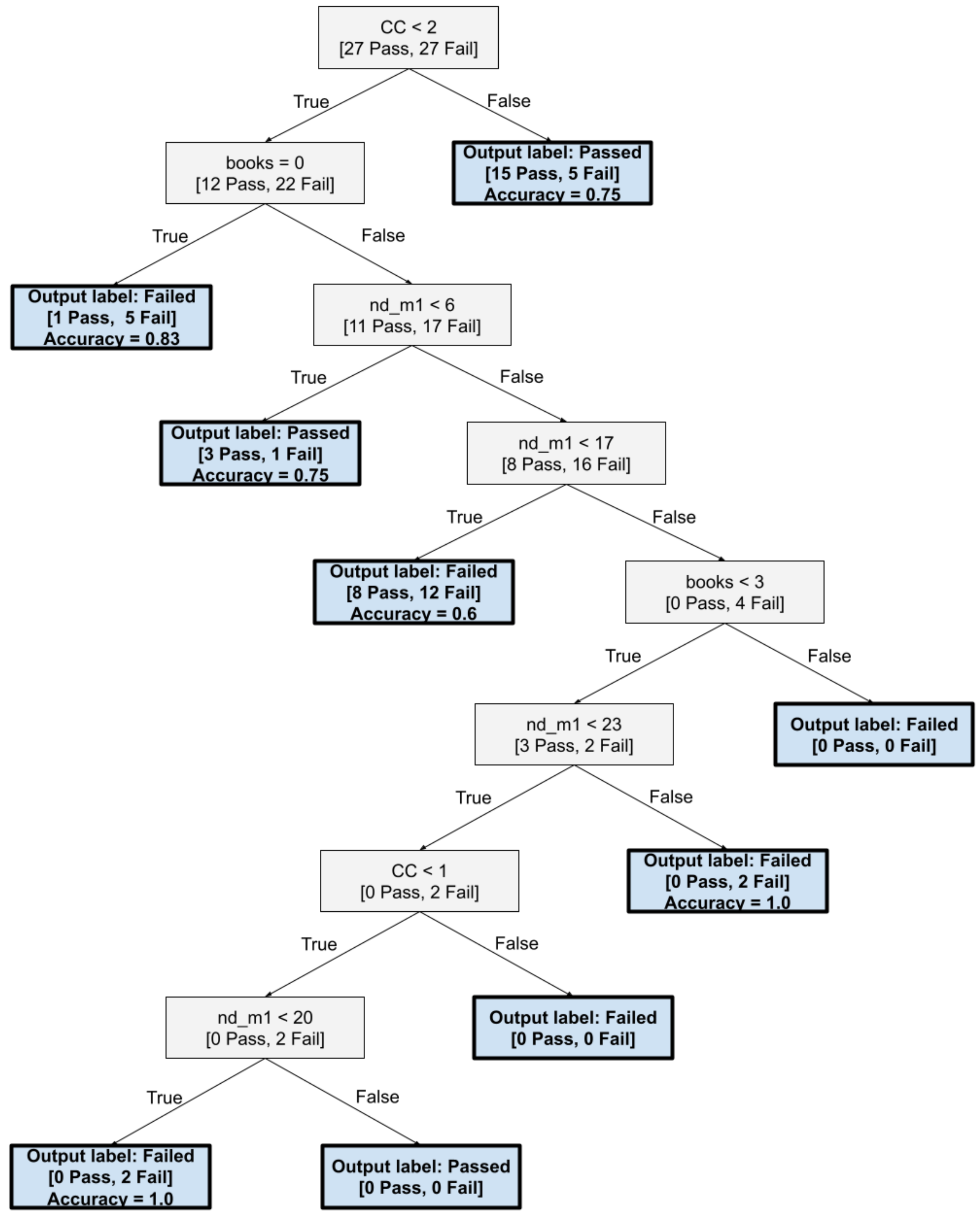

Figure 7. The decision Tree computed for the discipline 10/G1 and academic level AP for candidates with coverage A or B using the metrics *CC*, *books* and *nd_m1*.

### Analysis of Non-Citation-Based Metrics

Non-citation-based metrics also seem relevant in the evaluation process of both disciplines and academic levels. Specifically, *cand* has a high or above-average percentage of use at the FP level in both disciplines, and *books* has high or above-average percentage at the AP level in both disciplines. In 10/G1 AP, *co-au* has an above-average percentage as well. These results suggest that having published a book or having co-authored a publication with at least one member of the commission is a discriminating criterion in the evaluation of scholars applying to the role of AP. Whereas, candidates to the academic role of FP are evaluated based on their combined publishing efforts. It is also interesting to notice that our non-citation-based metrics appear to weigh more heavily in the assessment of candidates to 13/D4 than candidates to 10/G1. In 13/D4, not only do *cand* and *books* have higher percentages of use than in 10/G1, but also *co-au*, *books* and *other_pubbs* have high or above-average percentages at the FP level, while having very low values in 10/G1 FP.

**Analysis of Anvur Metrics**

ANVUR non-citation-based metrics are also relevant in the evaluation process of 13/D4. *ND_M3* (number of books), has an average percentage at the FP level. *ND_M1* (number of journal papers and book chapters) and *ND_M2* (number of papers published on *Class A* journals) respectively have a slightly above-average and a high percentage of use at the AP level. Consequently, these values combined with the ones described just above indicate that all aspects of a candidate's academic production are taken into consideration in the evaluation of 13/D4 applicants, especially at the FP level. In 13/D4 FP, *cand* maps the overall number of publications, *co-au* records publications co-authored with at least one member of the commission, *other_pubbs* tracks the publications that are not articles or books, and *books* and *ND_M3* measure the number of books in the CV and the overall number of books respectively. Similarly, in 13/D4 AP, *books* measure the number of books, *ND_M1* maps journal papers and book chapters and *ND_M2* tracks the number of *Class A* papers.

An additional interesting characteristic of 13/D4 is the very low number of classifiers with an F-measure above 0.700. This could indicate the great heterogeneity of the evaluation process for this specific discipline and role. Moreover, this could mean that the evaluation of the commission concerns aspects that are not captured by bibliometric indicators. Typical examples are the scientific responsibility of international projects, the editorship of international journals, academic awards and fellowships: all these examples attest the maturity of academics, but are inherently non-bibliometric aspects.

Candidates to the academic role of FP are mostly assessed on their good-standing within their scientific community and overall publishing efforts. Whereas, the evaluation of applicants to the academic role of AP heavily rests on whether a third party confirms their belonging to their scientific community, they authored a book, published a paper in a major journal, and co-authored a publication with the commission. Moreover, although the candidate's publication history is considered also in the evaluation of applicants to 10/G1, it is particularly important in the assessment of 13/D4 applicants.

**Comparing machine learning classifiers with statistical tests**

To further support our analysis of the relationship between our metrics and the result of the evaluation, we compare the 95% confidence intervals (CIs) of the bootstrapped means of each of our metrics between candidates that pass the evaluation and candidates that fail the evaluation. Specifically, we split our sample depending on the outcome of the evaluation, bootstrap the means of each bibliographic or citation-based metric, and calculate the confidence intervals of the distributions of means. Visually comparing the CIs provides us with insights regarding which metrics are similar between the candidates that pass and those that fail (and thus the distributions overlap), and which are different (therefore the distributions do not overlap).

Using this visual analysis, we notice that the CIs of bootstrapped means of candidates that fail the evaluation do not overlap with CIs of candidates that pass for the following citation-based metrics: *cand_comm* (97.5% CI Fail=0.90, 2.5% CI Pass=0.92), *comm_cand* (97.5% CI Fail=0.29, 2.5% CI Pass=0.43), *BC* (97.5% CI Fail=10.43, 2.5% CI Pass=14.65), and *CC* (97.5% CI Fail=1.10, 2.5% CI Pass=1.87).

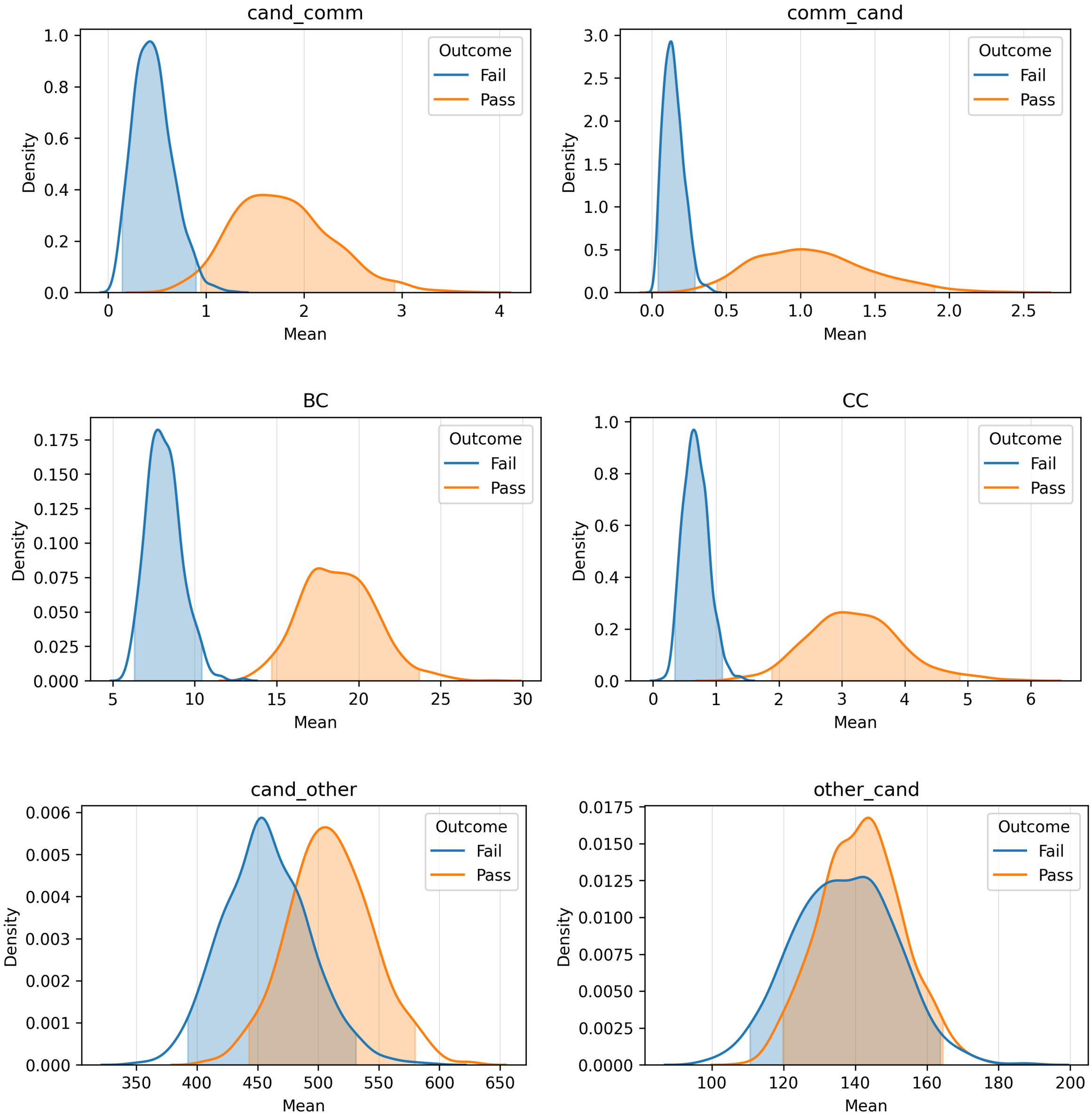

*Figure 8. Confidence intervals of the bootstrapped means for our citation-based metrics.*

The CIs overlap for all non-citation-based metrics. However, the shape of the two distributions for *co-au* is different than for the other non-citation-based metrics. It resembles more closely to the distributions of the means of the citation-based metrics for which we do not find an overlap. In addition, the CIs for *co-au* barely overlap (97.5% CI Fail=0.26, 2.5% CI Pass=0.24) compared to other metrics. To either support or dismiss that *co-au* values are different for candidates that pass the evaluation and those who fail, we conduct further statistical testing. The results of said tests are reported later in this section. Among ANVUR's metrics, the CIs do not overlap only in the case of *ND_M2* (97.5% CI Fail=5.78, 2.5% CI Pass=6.33).

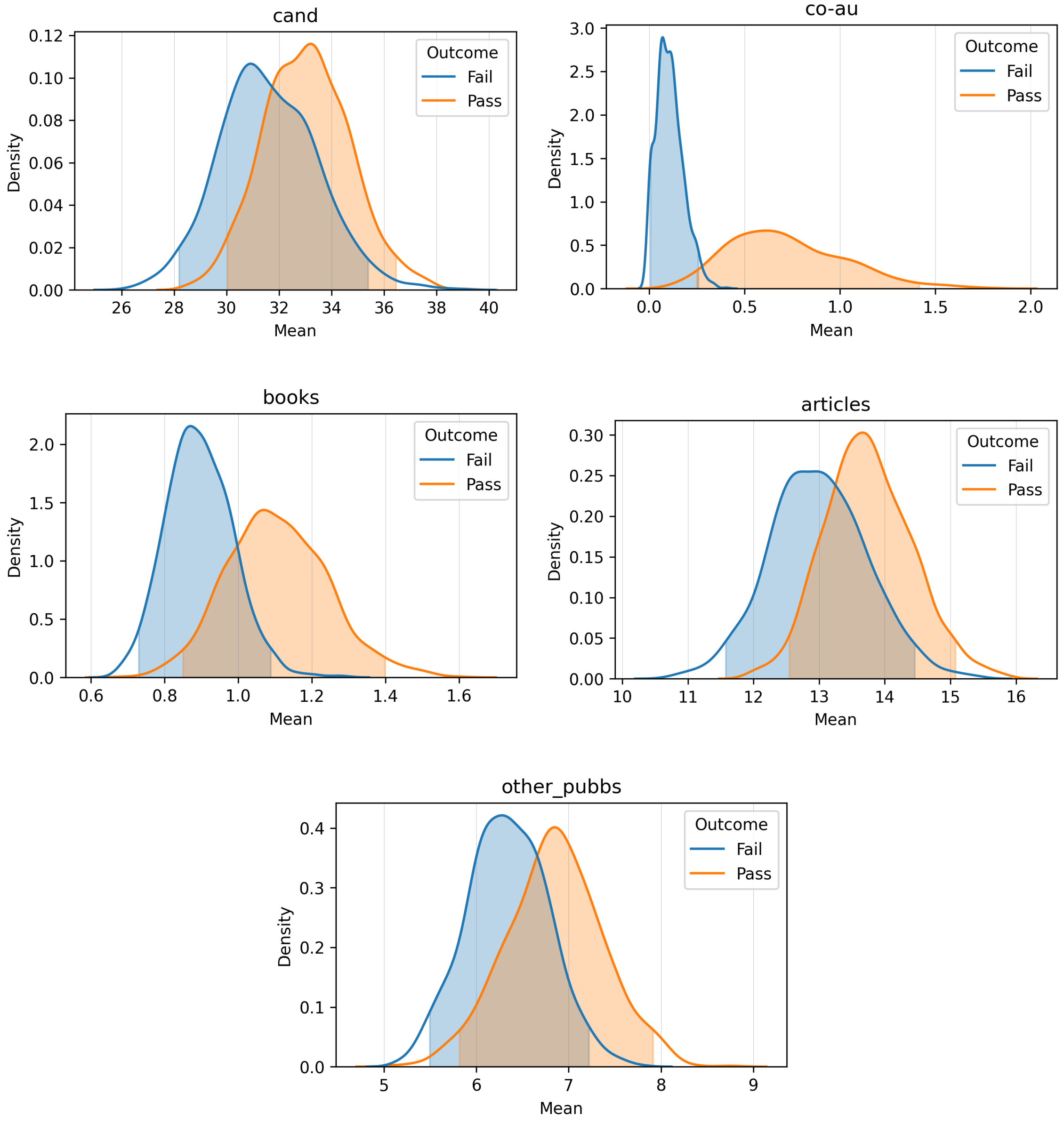

*Figure 9. Confidence intervals of the bootstrapped means for our non-citation-based metrics.*

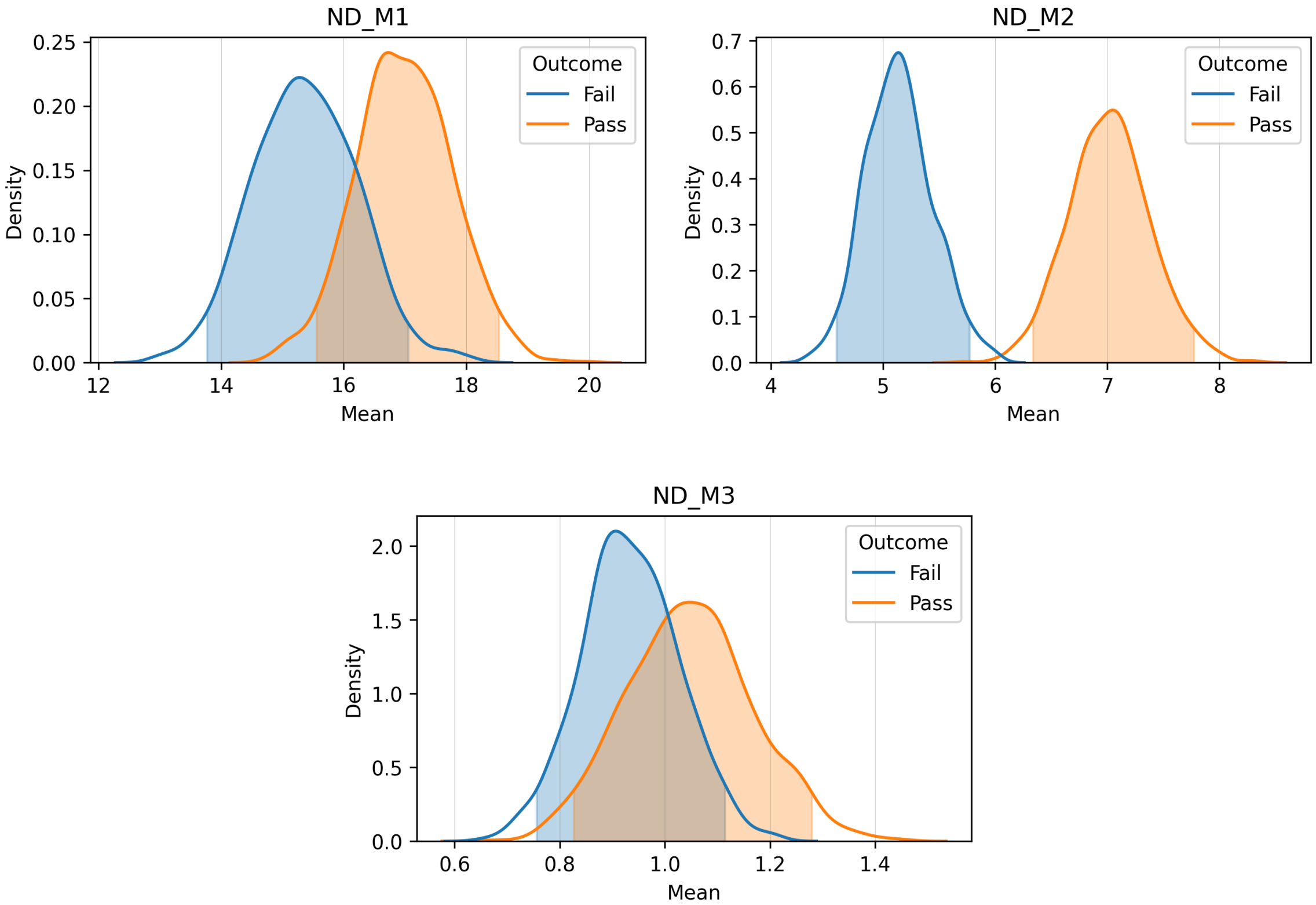

*Figure 10. Confidence intervals of the bootstrapped means for ANVUR's metrics.*

We carry out further statistical analyzes in order to confirm or dismiss that *cand_comm*, *comm_cand*, *BC*, *CC*, *ND_M2,* and potentially *co-au,* are statistically different between candidates that pass and candidates that fail the evaluation. Specifically, we carry out a permutation test and a Mann-Whitney U test for each of our metrics. Using these tests, we can determine whether there is a statistically significant difference between the sample of bibliographic and citation-based metrics of candidates that pass the evaluation and the sample of metrics of candidates that fail the evaluation. We choose these tests because they are both non-parametric and thus do not assume that the samples are normally distributed. Table 7 reports the metrics that are statistically different between our two samples (the candidates that passed the evaluation and the candidates that did not).

In line with our ML experiment's findings, we find that *cand_comm*, *comm_cand*, *BC*, *CC* are statistically different between the two samples. Candidates who pass the NSQ tend to have a higher number of citations to the commission and from the commission (*cand_comm* and *comm_cand*), and to share a higher number of citations with the commission (*BC* and *CC*). This further supports our findings regarding the importance of citation-based metrics for the evaluation of both disciplines and academic roles. However, these tests don't find a statistically meaningful difference between the two samples for *other_cand*, the number of citations from other scholars to the candidate. Of our non-citation-based metrics, the tests only identify *co-au* as statistically different between the two samples, as candidates who pass the NSQ have a higher number of publications they co-authored with at least one member of the commission. Whereas, our machine learning analysis also identifies the relevance of the number of publications (*cand*) and of books (*books*) authored by the candidate in the evaluation process. Among the ANVUR metrics, *ND_M2* is the only one that reaches statistical significance between the two samples. After all, *ND_M2* has the highest percentage among the ANVUR metrics in the results of our analysis.

| *Metric* | *Permutation* | *Mann-Whitney U* |
|---|---|---|
| **cand_comm** | Difference: -1.47<br>(p-value: 0.0) | *U* statistic: 26,915.0<br>(p-value: 7.43e-06) |
| **comm_cand** | Difference: -0.96<br>(p-value: 0.001) | *U* statistic: 27,580.5<br>(p-value: 1.90e-05) |
| **BC (bibliographic coupling)** | Difference: -11.86<br>(p-value: 0.0) | *U* statistic: 22,245.0<br>(p-value: 2.87e-08) |
| **CC (co-citation)** | Difference: -3.00<br>(p-value: 0.0) | *U* statistic: 23,652.5<br>(p-value: 1.34e-09) |
| **cand_other** | Difference: -57.44<br>(p-value: 0.102) | *U* statistic: 27,587.0<br>(p-value: 0.028) |
| **other_cand** | Difference: -5.69<br>(p-value: 0.372) | *U* statistic: 26,912.5<br>(p-value: 0.009) |
| **cand** | Difference: -2.35<br>(p-value: 0.166) | *U* statistic: 27,025.0<br>(p-value: 0.011) |
| **co-au** | Difference: -0.66<br>(p-value: 0.001) | *U* statistic: 28,821.0<br>(p-value: 0.001) |
| **books** | Difference: -0.15<br>(p-value: 0.16) | *U* statistic: 30,181.5<br>(p-value: 0.523) |
| **articles** | Difference: -0.74<br>(p-value: 0.245) | *U* statistic: 26,338.5<br>(p-value: 0.003) |
| **other_pubbs** | Difference: -0.76<br>(p-value: 0.141) | *U* statistic: 30,135.0<br>(p-value: 0.533) |
| **ND_M1** | Difference: -1.59<br>(p-value: 0.86) | *U* statistic: 23,054.5<br>(p-value: 0.012) |
| **ND_M2** | Difference: -1.88<br>(p-value: 0.0) | *U* statistic: 19,304.0<br>(p-value: 2.59e-07) |
| **ND_M3** | Difference: -0.11<br>(p-value: 0.246) | *U* statistic: 26,575.5<br>(p-value: 0.939) |

Table 7. Permutation test and Mann-Whitney *U* statistics and their associated p-values. For the permutation tests, the observed difference is calculated subtracting the mean of the sample of candidates who passed from the mean of the sample of candidates who did not pass.

**Looking at cognitive and social proximity in citations**

One of the aspects we discussed at the beginning of this paper is the mixture of cognitive and social aspects in the use of the citations. Citations are complex tools and an author can cite another one for multiple reasons, so that a citation can express different kinds of proximity: *cognitive* (common research interests, application of the same approach, common background, etc.) or *social* (same affiliation, same research group, past collaborations, etc.) or a combination of the two. Our work does not make any assumption on the nature of the citations and the indicators we calculated on the citation network. Though, it is possible to make some distinction. We consider BC (bibliographic coupling) and CC (co-citation) primarily as "cognitive connectors" between the candidates and the commissioners: the fact that a third party is involved - citing or being cited by both - makes it unlikely that the citation derives from a social connection only. The characterization of *cand_comm* and *cond_camm* indicators, on the other hand, is much less evident: the presence of a citation from/to the articles of candidates and commissioners might be the result of different forces, either cognitive and social.

Discerning between these forces is a very complex task, since there are no clear boundaries. We run a preliminary experiment on our dataset to start looking at these differences. This is not meant to be exhaustive - and we believe that such a distinction should be investigated in a dedicated article - but gives us a good starting point for future research.

In particular, we studied the co-affiliation network of the candidates and commissioners in our dataset, with the objective of measuring the correlation between the number of years a candidate worked in the same institution of at least one commissioner and the parameters *cand_comm* and *cond_camm* indicators; we also studied the correlation between the co-affiliation and the final outcome in the NSQ. The affiliations were taken from "Cerca

Università" (https://cercauniversita.cineca.it/) a publicly available service to retrieve information about the Italian Universities, including data about departments, people and courses. The service contains one record for each year in which a given person was affiliated to an Italian University. The same person can then have multiple records, with different affiliations, for her/his career.

Table 8 summarizes the data we were able to collect. We considered all 500 applications of our experiment and 11 commissioners. For 352 applications (about 70%) we were able to find at least one affiliation in an Italian University. The other candidates are missing for two reasons. First of all, it might happen that a candidate applied for the NSQ even if she/he works abroad; in that case there is no record in "Cerca Università" but we can safely conclude that the candidate and the commission member have never worked in the same institution. The second case is a bit more tricky: "Cerca Università" only records people with a tenure track or a permanent position; phD students, postdocs and adjunct professors are then not listed and considered in our analysis. This is indeed a limitation of the experiment. Nonetheless, the connection between these candidates and the commissioners is weaker than the others: a PhD student, for instance, might work in the same department of a commissioner even without knowing each other; the same might apply for adjunct professors.
A complete analysis of social proximity should also consider these cases but, at this stage, we only focus on permanent positions and co-affiliations and consider 352 candidates for which we have reliable data.

| **500** unique applications |
|---|
| **11** commissioners (all with Italian affiliation) |
| ***352*** **applications of candidates with Italian affiliations** |
| **33 candidates/commissioners in the same department (0.09%)** |
| **53 candidates/commissioners in the same University (0.14%)** |

Table 8. Data about co-affiliations.

We then searched for co-affiliations, counting the number of years in which each candidate worked in the same institution of a commission member. We considered two cases: (a) the exact same department and (b) different departments but in the same University.

We found that only 33 candidates were affiliated to the same department of at least one commission member (9%), and the number increases to 52 if we consider co-affiliation at University level (14%). These data, though limited, indicate that the influence of such a proximity is limited.

We then measured the percentage of these candidates who achieved a positive outcome in the NSQ, so as to compare that percentage to the same value over the full dataset. As reported before, in total, 48,3% of candidates passed the NSQ in the 10/G1 field and 48,7% in 13/D4. The result is interesting: these percentages increase significantly when restricting the analysis to the candidates with co-affiliations only. In fact, for 10/G1 the percentage increases up to 66,66% (same department) and 65,21% (same university) against the overall 48,3%. Similar figures were found for the 13/D4 field: the percentage goes up to 77,77% (same department) and 65,52% (same university) against 48,7% (overall).

We also investigated whether the aforementioned co-affiliation indices are related to the results of the NSQ. The approach to answer this question is to compare the NSQ successful and unsuccessful candidates through co-affiliation indices. Since the co-affiliation indices do not follow a normal distribution (as shown in Table 9 and Figure 11), we chose the Wilcoxon–Mann–Whitney test applying the continuity correction in the normal approximation for the *p*-value.

| **Index** | **Shapiro-Wilk test (W)** | **p-value** |
|---|---|---|
| years_co-affiliation_university – successful candidates | 0.340 | $1.193e^{-25}$ |
| years_co-affiliation_university – unsuccessful candidates | 0.193 | $3.542e^{-26}$ |
| years_co-affiliation_department – successful candidates | 0.475 | $2.591e^{-23}$ |
| years_co-affiliation_department – unsuccessful candidates | 0.302 | $1.105e^{-24}$ |

**Table 9:** Normality Assessment of co-affiliation indices

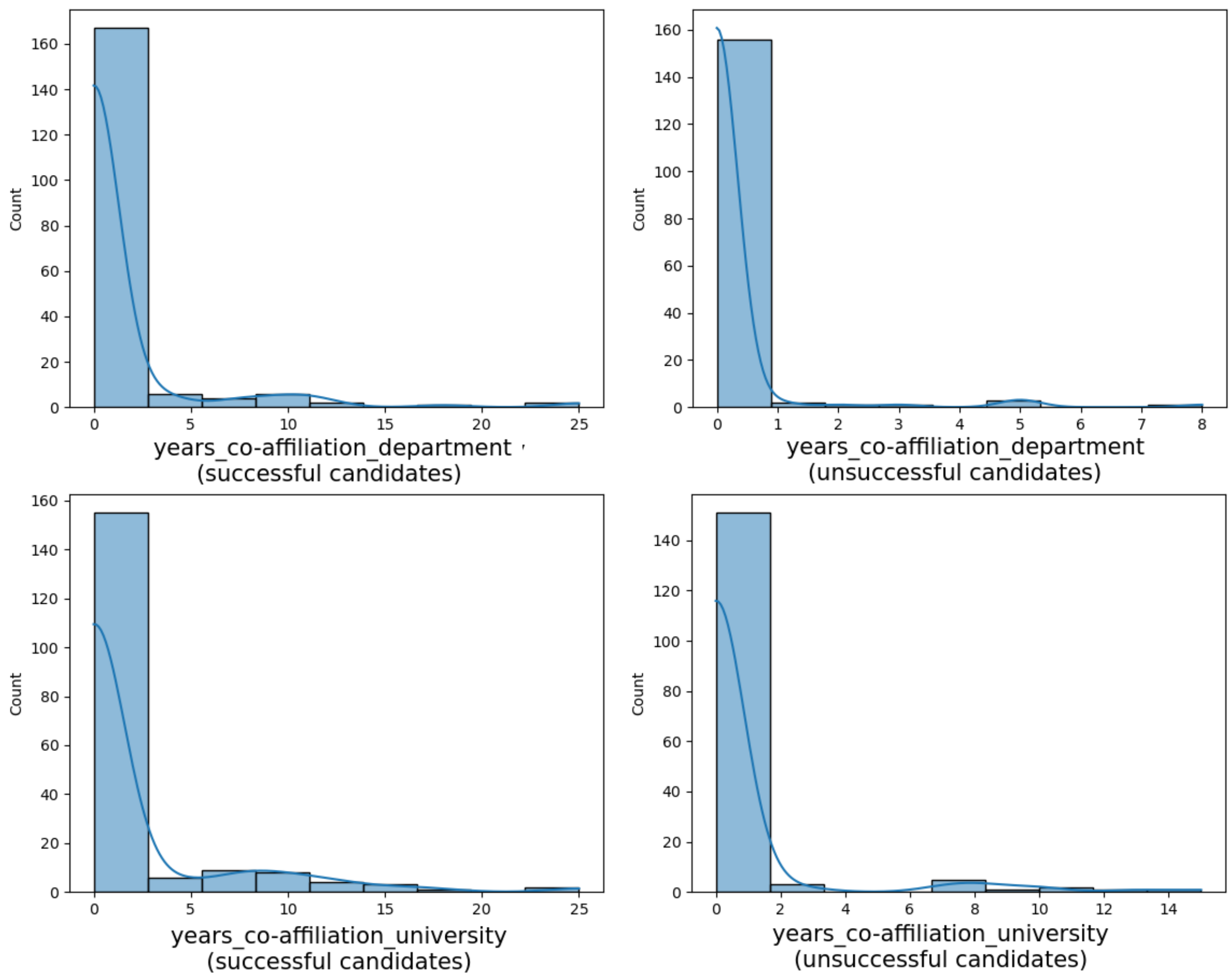

**Figure 11:** histograms of the two co-affiliation indices for successful and unsuccessful NSQ candidates.

The analysis was carried out on all candidates, regardless of discipline, because the aim of this step is to assess whether the co-affiliation indices differ significantly between the two groups, showing them to be a priori potential explanatory factors for the outcome of the NSQ. Table 10 shows the results of Wilcoxon–Mann–Whitney test performed on each co-affiliation. Both indices tested were significantly different in the two groups, with averages always higher in the successful candidates' group. These results support the idea of using co-affiliation and other indices measuring the social proximity as explanatory variables of NSQ.

| Index | Successful candidates mean | Unsuccessful candidates mean | W | p |
|---|---|---|---|---|
| years_co-affiliation_university | 1.106 | 0.183 | 16680 | $7.803e10^{-3}$ |
| years_co-affiliation_department | 1.809 | 0.646 | 17022 | $7.652e10^{-3}$ |

**Table 10:** Wilcoxon–Mann–Whitney tests for the co-affiliation indices between the NSQ successful and unsuccessful candidates groups. Indices are computed using the data from all candidates regardless of discipline.

This seems to imply that the co-affiliation relates to the outcome of the evaluation, even if it is not considered explicitly. More experiments are needed to confirm this hypothesis, considering the small amount of data used for this experiment. Note also that other social connections could be in place - for instance, the participation in the same research project or consortium or the co-organization of events or co-editing of journals - and are worth being included in a more exhaustive analysis of social proximity that we plan to integrate with the current work.

**Drawing conclusions and future directions**

The results of our experiments with ML classifiers draw two orthogonal conclusions. On the one hand, they seem to highlight the existence of a possible bias between applicants to the NSQ and the related committee members observed by analyzing the citational relationships between the researchers, even in those contexts where such citations are not explicitly taken into account into the NSQ process. Indeed, we found that the connections between the citation networks of the candidates and the commissioners are relevant for the peer-review process even for NDs disciplines.

We believe that this is necessarily not a direct causal-effect connection with a negative connotation - i.e. a candidate is supported by a commission member because they have some proximity, being cognitive or social. Rather, this is a consequence of the use of the citations: commissioners are well-known in the community and recognized experts in their field; citing their works - or being cited by these works - also implies being well-connected to the same community and experts in that area.

The second conclusion is about the ML process we took in place. The fact that the evaluation process can be studied by looking at the results of machine learning techniques suggests new applications for these techniques: not just predicting but also analyzing *a posteriori* the human-based evaluation. In fact, we do not advocate the adoption of machine learning and other artificial intelligence technologies as a mean to replace human judgement in research assessment exercises. Rather, we believe that these technologies enable looking and analysing the current assessment practices from a new perspective by highlighing possible hidden (conscious and unconscious) relations among the different actors involved in the process. For instance, such tools might be used for scanning a process (such as a research assessment exercise) to identify emerging issues within it - for instance influenced by proximity - thus enabling humans to look closely at these issues and, eventually, to find workarounds to mitigate them and improve the process itself.

In future work, we will expand this study by taking into consideration all NDs, and including all CDs as well. Indeed, we intend to continue this investigation by proceeding into two separate, yet related, research directions. On one hand, we will analyze the coverage of both CDs and NDs by open datasets. We will draw comparisons across different NDs and between CDs and NDs. This would give us the opportunity to analyze more thoroughly how coverage of the candidates' publications vary across disciplines and research topics. On the other hand, we plan to study the relevance of our bibliographic and citation-based metrics in the qualitative evaluation of NDs and CDs. Thus, we would be able to determine whether citation-based metrics have a larger impact on the qualitative peer-review of CDs in contrast to NDs.

In addition, we also plan to extend the analysis of social proximity we started in this work, by further investigating how cognitive and social proximity correlate on NSQ data, as done by (Boyack & Klavans, 2010) and (Yan & Ding, 2012), who provided very accurate comparisons between these social indicators as tools for research trend analysis and evaluation. We will also investigate community detection techniques, like the Leiden algorithm (Traag et al., 2019) that proved to outperform the well-known Louvain algorithm both in terms of performance and reliable identification of connected (sub-)communities, as well as metrics to measure network centrality (Stephenson & Marvin, 1989) and their application to research collaborations measurement (Zhan et al., 2014).

# Declarations

### Funding and other competing interests

The authors did not receive support from any organization for the submitted work. The authors have no relevant financial or non-financial interests to disclose.